\documentclass[journal,onecolumn,romanappendices]{IEEEtran}
\usepackage[english]{babel}
\addto\captionsenglish{%
}
\usepackage{amsmath,amsfonts, amssymb, latexsym}
\usepackage{macros}
\usepackage[ruled,norelsize]{algorithm2e}
\usepackage{dsfont}
\usepackage{bbold}
\usepackage{cite}
\usepackage[caption=false, font=footnotesize]{subfig}
\usepackage{graphicx}
\usepackage{tikz,xifthen}
\usepackage{url}
\usepackage{comment}
\usepackage{fix-cm}    
\makeatletter
\newcommand\HUGE{\@setfontsize\Huge{20}{30}}
\makeatother   
\DeclareGraphicsExtensions{.pdf,.jpeg,.png,.jpg,.eps}

\allowdisplaybreaks

\newtheorem{theorem}{Theorem}
\newtheorem{lemma}{Lemma}
\newtheorem{corollary}{Corollary}
\newtheorem{proposition}{Proposition}

\ifCLASSINFOpdf
\else
\fi
\begin{document}

\title{Fundamental Limits of Noncoherent Massive Random Access Networks}

\author{Grace Villacr\'es,~\IEEEmembership{Member,~IEEE,}
        Tobias Koch,~\IEEEmembership{Senior Member,~IEEE,}
        and~Gonzalo Vazquez-Vilar,~\IEEEmembership{Senior Member,~IEEE}
        \thanks{G.~Villacr\'es has been partially supported by Comunidad de Madrid within the 2023-2026 agreement with Universidad Rey Juan Carlos for the granting of direct subsidies for the promotion, encouragement of research and technology transfer, line of Action A Emerging Doctors, under Project OrdeNGN (Ref. F1177). T.~Koch and G.~Vazquez-Vilar have received funding from the European Research Council (ERC) under the European Union's Horizon 2020 research and innovation programme (Grant No.~714161), from the Spanish Ministerio de Ciencia e Innovación under Grants RYC-2014-16332, TEC2016-78434-C3-3-R (AEI/FEDER, EU),  PID2020-116683GB-C21~/~AEI~/~10.13039/501100011033, and PID2024-159557OB-C21~/~AEI~/~10.13039/501100011033, and from the Comunidad de Madrid under Grant IDEA-CM (TEC-2024/COM-89). The material in this paper was presented in part at the 2016 IEEE International Symposium on Information Theory, Barcelona, Spain, July 2016 \cite{Villacres16} and in part at the 2020 IEEE International Symposium on Information Theory, Los Angeles, CA, USA, June 2020 \cite{Villacres20}.}
\thanks{G. Villacr\'es is with the Department
of Signal Theory and Communications, Universidad Rey Juan Carlos, 28942 Fuenlabrada,  Spain (e-mail: grace.villacres@urjc.es).}
\thanks{T. Koch and G. Vazquez-Vilar are with the Department
of Signal Theory and Communications, Universidad Carlos III de Madrid, 28911 Leganés,  Spain, and also with the
Gregorio Marañón Health Research Institute, 28007 Madrid, Spain (emails: tkoch@ing.uc3m.es, gonzalo.vazquez@uc3m.es).}}
\maketitle


\begin{abstract}
This paper studies the capacity of massive random-access cellular networks, modeled as a multiple-input multiple-output fading channel with an infinite number of interfering cells. To characterize the symmetric sum rate of the network, a random-coding argument is invoked together with the assumption that in all cells users draw their codebooks according to the same distribution. This assumption can be viewed as a generalization of the assumption of Gaussian codebooks, often encountered in the literature. The network is further assumed to be noncoherent: the transmitters and receivers are cognizant of the statistics of the fading coefficients, but are ignorant of their realizations. Finally, it is assumed that the users access the network at random with a given activation probability. For the considered channel model, rigorous upper and lower bounds on the capacity are derived. The behavior of these bounds depends critically on the path loss from signals transmitted in interfering cells to the intended cell. In particular, if the fading coefficients of the interferers (ordered according to their distance to the receiver) decay exponentially or more slowly, then the capacity is bounded in the transmit power. It thus confirms that the saturation regime in interference-limited networks---observed by Lozano, Heath, and Andrews (``Fundamental limits of cooperation", \emph{IEEE Trans.\ Inf.\ Theory}, Sept.~2013) for the more restrictive case where users are always active and the channel inputs are restricted to a scale family---cannot be avoided by random user activity or by using channel inputs beyond the scale family. In contrast, if the fading coefficients of the interferers decay faster than double-exponentially, then the capacity is unbounded in the transmit power. Note that proving an unbounded capacity is nontrivial even if the number of interfering cells is finite, since the condition that the users’ codebooks follow the same distribution prevents interference-avoiding strategies such as time-, frequency-, or code-division multiple access. We obtain this result by using bursty signaling together with treating interference as noise.
\end{abstract}


%
\IEEEpeerreviewmaketitle

\section{Introduction}
\IEEEPARstart{T}{he} rapid evolution of modern and future wireless communication technologies is driven by the growing demand for enhanced data rates, extended coverage, and improved scalability. Emerging applications, such as high-speed mobile broadband, the \emph{Internet of Things (IoT)}, and \emph{machine-type communications (MTC)}, require networks to support both human users and a vast number of interconnected devices. As a result, technological advancements are not only improving user experiences but also addressing the increasing complexity of managing heterogeneous devices and services \cite{Fettweis21, Akyildiz20, Alsabah21}.

A key factor in this transformation is the exponential growth of IoT devices. According to the IoT Analytics Report \cite{IoTAnalytics24}, the number of connected IoT devices is expected to reach 40 billion by 2030. This growth, coupled with the rising number of network users, is expected to significantly increase data traffic and place higher demands on data rates. To meet these demands, future wireless networks will rely heavily on radio access network densification. However, this densification also introduces new challenges, particularly in handling traffic bursts. These bursts, often caused by sporadic user activity or the unpredictable nature of IoT and MTC devices, can overload network resources and degrade performance if not properly managed \cite{Kumar24}.

\begin{figure}
    \centering
    \subfloat[eMBB: cell-based network.]{
        \includegraphics[width=.48\linewidth]{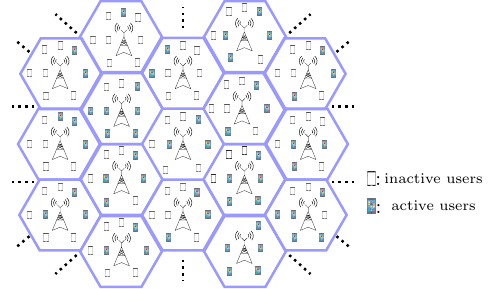}
        \label{fig:cell-based}}
        \hfill
        \subfloat[mMTC: cell-free based network.]{
        \includegraphics[width=.35\linewidth]{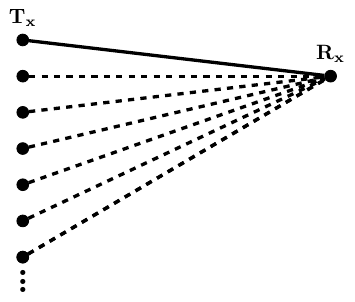}
        \label{fig:grant-free}}
    \caption{Comparison of eMBB and mMTC network architectures.}
    \label{fig:combined}
\end{figure}

In this context, wireless communication technologies are evolving along two main paradigms:

\emph{Enhanced mobile broadband (eMBB)} supports applications that require high data rates and depend on dense network deployments and advanced interference management. These systems are typically deployed using a cell-based network architecture (see Figure \ref{fig:cell-based}) and use, for example, femtocells and macrocells to share network resources effectively. However, this densification, while enhancing capacity, also increases the risk of inter-cell interference as the number of communicating users grows. Since interference is one of the main limiting factors for achieving higher data rates, its management has been the subject of several studies; see, e.g., \cite{ZahirFemtocells} and references therein.

\emph{Massive machine-type communications (mMTC)} in turn, is tailored for massive connectivity, focusing on supporting a large number of devices with sporadic activity, as seen in IoT systems. This approach typically employs a cell-free network architecture, allowing devices to communicate directly with multiple access points (see Figure \ref{fig:grant-free} for the case of only one access point). Access to the network is often based on grant-free random access, enabling devices to transmit data without waiting for scheduling or resource allocation \cite{Choi21, Akyildiz20, Ozates24}. In scenarios involving hundreds of millions of potentially connected devices, Polyansky \cite{Polyanskiy17} introduced the concept of \emph{unsourced random access (URA)}. In URA, all users share a common codebook, and the active users select codewords from it to transmit their messages. The receiver focuses on decoding the content of the messages without needing to identify the individual users, thereby reducing signaling overhead. In this scenario, inter-user interference also becomes a critical challenge due to the large number of potentially transmitting devices.

\subsection{Research Scope and Problem Statement}

There are several lines of research that have studied different aspects of wireless networks. For example, the impact of interference in wireless networks was investigated under the framework of the \emph{interference channel}; see, e.g., \cite{elgamal2011book, Etkin, cadambe_interference_2008}. While, traditionally, it was assumed that in these networks interference is always present, more recent works have also considered intermittent/bursty interference due to intermittent user activity (as in IoT networks) or opportunistic frequency reuse among cells \cite{Hamdaoui2009}; see, e.g., \cite{Khude09, Wang13, Villacres18}. More recently, the problem of a massive number of users accessing sporadically a wireless network was studied under the frameworks of the \emph{many-access channel} \cite{ChenCG17} or \emph{massive random access} \cite{Polyanskiy17}. In the many-access channel, the number of users accessing the network grows to infinity with the blocklength to model a massive number of users.
This problem has attracted significant attention in the information theory literature, with contributions ranging from achievability and converse bounds for Gaussian channels \cite{ZadikPT19,Ravi22,Ngo23,LiuISIT24,Liu24} and fading channels \cite{KowshikP19,KowshikTCOM20,Gao25} to the development of practical coding schemes for this setting \cite{OrdentlichP17,Vem19,Fengler19,Amalladinne20,Fengler21,Amalladinne22,Hsieh22,LiuISIT24,Liu24}.
When the number of users grows without bound alongside the blocklength $n$, the number of bits each user can reliably transmit scales sublinearly in $n$~\cite{ChenCG17}. In fact, many of the aforementioned works assume that each user transmits a fixed number of bits. This implies that the number of bits that each user can transmit per channel use vanishes as the blocklength tends to infinity. 

A different approach to studying the fundamental limits of wireless networks was followed by Lozano, Heath, and Andrews \cite{lozano_fundamental_2013}, who modeled the wireless network as a multiple-input multiple-output (MIMO) block-fading channel and then studied its capacity. They showed that the maximal coding rate achievable with channel inputs of the form $\sqrt{\mathsf{P}} X$ (where $X$ is normalized, so that $\mathsf{P}$ represents the transmit power, and the distribution of $X$ does not depend on $\mathsf{P}$) is bounded in the transmit power $\mathsf{P}$. This suggests that there is a saturation regime where the channel capacity hits a ceiling that is independent of the transmit. However, while Gaussian inputs can be written in the above form, limiting the analysis to such inputs may be overly restrictive and could potentially be the reason for a bounded capacity.

In this paper, we follow the approach in \cite{lozano_fundamental_2013} and model the wireless network as a MIMO fading channel. We assume that there is an infinite number of interferers to model a massive number of users. More precisely, we model a network where $n_T$ users are grouped inside a cell, each communicating with a base station equipped with $n_R$ antennas. Users within the same cell are allowed to cooperate, eliminating intra-cell interference, while inter-cell interference arises from users in neighboring cells. In contrast to the vast literature on massive random access, we assume that there is an infinite number of interferers (rather than a number of interferers that grows with the blocklength), but we account for path loss to ensure that the interference from distant users diminishes, thereby permitting the transmission at a positive coding rate. Furthermore, in contrast to \cite{lozano_fundamental_2013}, we consider intermittent user activity, i.e., users access the network at random with a given activity probability. We further lift the restriction on the channel inputs imposed in \cite{lozano_fundamental_2013}. We believe this model captures the essence of both eMBB and mMTC scenarios. Specifically, the eMBB scenario corresponds to the above model for general $n_T$ and $n_R$ (see Figure~\ref{fig:cell-based}), while the mMTC scenario can be modeled by setting $n_T=n_R=1$ (see Figure~\ref{fig:grant-free}). In contrast to an earlier version of this work \cite{Villacres20}, we treat interference at a per-user level rather than a per-cell level, thus capturing the impact of user activity more accurately. This model better approximates new-generation wireless communication systems, which are expected to support emerging IoT and \emph{machine-to-machine (M2M)} communication services, typically characterized as mentioned before by a massive number of IoT devices with different types of data traffic and sporadic, bursty activity of the individual users.

\subsection{Contributions and Organization}

We analyze the behavior of channel capacity in noncoherent wireless networks at high transmit power, focusing on the scenario where users and receivers only know the statistics of the fading coefficients, but not their realizations---an assumption that we believe is realistic when the number of interferers is large. We establish explicit conditions under which the capacity remains bounded in the transmit power, revealing a severe power inefficiency in these networks. Our key findings in this work can be summarized as follows:

\begin{enumerate}
    \item \textit{Bounded Capacity in Noncoherent Wireless Networks:} 
    We rigorously establish an upper bound on the channel capacity, showing that capacity remains finite under practical fading conditions. Specifically, we prove that if the variances of the fading coefficients of the interferers decay at an exponential rate or slower, then the channel capacity is bounded in the transmit power. Thus, the presence of interference fundamentally limits the achievable rate, regardless of how high the transmission power is. This insight is validated through well-known propagation models, including the free-space path loss model, the two-ray model, and the Okumura-Hata model.

    \item \textit{Unbounded Capacity under Faster-than-Double-Exponential Fading Decay:} 
    We establish that, if the variances of the fading coefficients of the interferers decay faster than double-exponentially, then the channel capacity is unbounded in the transmit power. This result highlights the critical role of the decay rate of fading variances in determining fundamental limits on information transmission.

    \item \textit{Effect of Interference Burstiness:} 
    To restore the unbounded capacity growth for faster than double-exponentially fading decays, we propose an intermittent signaling scheme that artificially induces interference burstiness, creating opportunities where the channel is effectively interference-free. We therefore show the impact of interference burstiness on the network capacity, and demonstrate that under a sufficiently sparse user activity pattern, unbounded capacity growth can be restored.

    \item \textit{Extension to General Noncoherent Wireless Scenarios:} 
    The proof techniques used in this work rely on aligning interfering signals across multiple cells in an eMBB scenario or across multiple users in an mMTC scenario. By exploiting the fact that all users draw their codebooks from the same distribution, we partially cancel interference effects, offering a new analytical framework that may be extended to other communication settings.
\end{enumerate}

The rest of this paper is organized as follows. Section~\ref{sec:model} introduces the cellular network model, detailing the system setup, key assumptions, and the fading model considered in our analysis. In Section~\ref{sec:bounded-capacity}, we rigorously establish an upper bound on the channel capacity, demonstrating that under practical fading conditions, capacity remains bounded in the transmit power. We derive explicit conditions under which this result holds and validate it through representative propagation models. Section~\ref{sec:bursty-signaling} shifts focus to achievable rates, showing that bursty signaling schemes can effectively mitigate interference and, under specific conditions, enable unbounded capacity growth. Finally, Section~\ref{sec:discussion} concludes the paper with a summary and discussion of our results. Proofs and technical derivations are deferred to the appendices.

\begin{figure}
    \centering
    \includegraphics[width=0.5\linewidth]{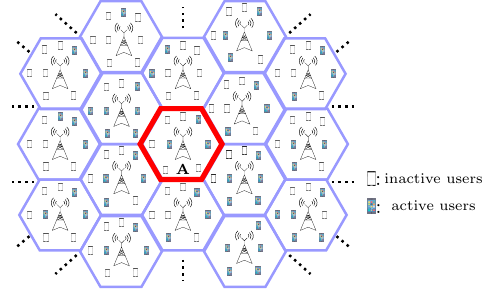}
      \caption{Cellular network model.}\label{fig:cellular-model}
\end{figure}

\section{Network Model} \label{sec:model}

We consider a cellular network in which users are grouped inside cells and communicate with a base station. Users inside each cell are assumed to cooperate, hence, there is no intra-cell interference. However, these users do not cooperate with the users in other cells. Therefore, the transmissions from other cells interfere with this communication. The activity of the users within the network is assumed to be intermittent, so only a set of users is active at the same time. Figure~\ref{fig:cellular-model} illustrates the network model. Cell \textbf{A} (in red) corresponds to the intended cell, and there are an infinite number of interfering cells. In each cell, only a fraction of the users is active at the same time. 

Our performance measure is the capacity of the channel between the users and the base-station inside the intended cell (uplink transmission). Since a characterization of all achievable rates in the network is unfeasible when the number of cells and users is large, we study the \emph{symmetric sum rate} of the network, i.e., the sum rate that can be achieved by users within the intended cell when all cells communicate at the same sum rate. Additionally, we consider the following assumptions:
\begin{enumerate}
\item[A1.] \textit{Infinitely Many Interfering Cells:} To model a large network, we assume that there are infinitely many interfering cells. This assumption does not restrict the generality of the problem, as interfering cells are parametrized by fading variances $\alpha_{\ell}$, $\ell=1,2,\ldots$, which can be set to $\alpha_{\ell}=0$ for inactive cells.
\item[A2.] \textit{No Cooperation:} We assume that users from different cells do not cooperate. This restriction excludes the use of coordination strategies such as \emph{time-division multiple access (TDMA)} or \emph{frequency-division multiple access (FDMA)}, which could otherwise enable unbounded capacity in specific time or frequency slots.
\item[A3.] \textit{Uniform Coding Across Cells:} To ensure that all cells (intended and interfering) communicate at the same rate, we invoke a random-coding argument and assume that all users within a cell draw their codebooks from the same distribution. This assumption can be viewed as a generalization of the assumption of Gaussian codebooks, which is often encountered in the literature. It is also reminiscent of the URA approach \cite{Polyanskiy17} in massive random access scenarios, where nodes use the same codebook.
\end{enumerate}

\subsection{Channel Model and User Activity}

Let us denote by $n_T$ the number of users located in a cell, and by $n_R$ the number of receive antennas within the base station. The channel input-output relation at time $k \in \Integers$  is given by
\begin{equation}\label{eqn:channel-model}
\vect{Y}_{k}= \vmat{H}_{0,k}\vect{X}_{0,k}+\sum_{\ell =1}^{\infty} \vmat{H}_{\ell,k}\vect{X}_{\ell,k} + \vect{Z}_k,
\end{equation}
where
\begin{itemize}
\item  $\vmat{H}_{0,k} \in \mathbb{C}^{n_R\times n_T}$ denotes the matrix of fading coefficients of the links from the transmitters inside the intended cell;
\item $\vmat{H}_{\ell,k} \in \mathbb{C}^{n_R \times n_T}$ denotes the matrix of fading coefficients of the interfering links from the transmitters inside the $\ell$-th interfering cell;
\item $\displaystyle{\vect{X}_{0,k}=[X_{0,1,k},\ldots,X_{0,n_T,k}]^T\in\mathbb{C}^{n_T\times 1}}$ corresponds to the vector of symbols transmitted at time $k$ by users inside the intended cell;
\item \mbox{$\vect{X}_{\ell,k}=[X_{\ell,1,k},\ldots,X_{\ell,n_T,k}]^T\in\mathbb{C}^{n_T\times 1}$} denotes the vector of symbols transmitted at time $k$ by the users inside the $\ell$-th interfering cell;
\item $\vect{Z}_k \in \mathbb{C}^{n_R\times1}$ models the \mbox{time-$k$} additive noise vector.
\end{itemize}
As is common, we assume that the additive noise $\{\vect{Z}_k,\,k\in\mathbb{Z}\}$ is a sequence of \emph{independent and identically distributed (i.i.d.)} random variables with a circularly-symmetric, complex Gaussian distribution of zero mean and covariance matrix $\sigma^2\mathsf{I}$, i.e., $\{\vect{Z}_k,\,k\in\mathbb{Z}\}\sim \textnormal{i.i.d.\ } \mathcal{N_{\mathbb{C}}}(0,\sigma^2\mathsf{I})$, where $\mathsf{I}$ denotes the identity matrix and $\mathcal{N_{\mathbb{C}}}(\mu,\mathsf{K})$ denotes the circularly-symmetric, complex Gaussian distribution with mean $\mu$ and covariance matrix $\mathsf{K}$. For simplicity, we further assume that the fading coefficients $\{\vmat{H}_{0,k},\,k\in\mathbb{Z}\}$ and $\{\vmat{H}_{\ell,k},\,k\in\mathbb{Z}\}$, $\ell=1,\ldots$ are sequences of i.i.d.\ random matrices, the former with i.i.d.\ $\mathcal{N_{\mathbb{C}}}(0,1)$ entries and the latter with i.i.d.\ $\mathcal{N_{\mathbb{C}}}(0,\alpha_{\ell})$ entries for some $\alpha_{\ell}$, $\ell=1,2,\ldots$ 
We consider a noncoherent scenario where transmitter and receiver only know the statistics of the fading coefficients but not their realizations.

The locations of the interfering cells relative to the intended cell enter the channel model through the variances $\alpha_{\ell}$, $\ell=1,2,\ldots$ of the fading coefficients $\{\vmat{H}_{\ell,k},\,k\in\mathbb{Z}\}$, $\ell=1,\ldots$ For simplicity, we assume that the entries of $\vmat{H}_{\ell,k}$ have identical variances. This corresponds to the case where all nodes in the $\ell$-th interfering cell are at the same distance from the receiver of the intended cell. This assumption can be relaxed to different variances, provided that they are proportional to $\alpha_{\ell}$. We assume that the total power of the interference received at the intended cell is finite, i.e.,
\begin{equation}
\sum_{\ell=1}^{\infty} \alpha_{\ell}<\infty.
\end{equation}
Without loss of generality, we order the interfering cells according to the variances of the corresponding fading variances, i.e., $\alpha_{\ell}\geq\alpha_{\ell'}$ for any $\ell<\ell'$. See also Assumption~A1.

The activity of the users both inside the intended cell and the interfering cells is assumed to be intermittent. We model this activity as
\begin{equation}\label{eqn:activity-model}
  X_{\ell,u,k} = {B}_{\ell,u}\tilde{X}_{\ell,u,k},\quad \ell=0,1,\ldots
\end{equation}
where $\tilde{X}_{\ell,u,k}$ denotes the symbol transmitted at time $k$ by user $u$ within cell $\ell$, and where ${B}_{\ell,u}$  is a random variable that captures the activity of this user. We shall model this activity by a Bernoulli random variable ${B}_{\ell,u}\sim\text{Ber}(\delta)$, which remains constant during the entire transmission, i.e., for $k=1,2,\ldots,n$. Thus, users are active in bursts with probability $\delta$ for some $0\leq\delta\leq 1$. We further assume that the user activities $B_{\ell,u}$ of different users (within the same cell or in different cells) are independent, i.e., the random variables $B_{\ell,u}$, $u=1,\ldots,n_T$, $\ell=0,1,\ldots$ are i.i.d.

Let $\tilde{\vect{X}}_{\ell,k}=[\tilde{X}_{\ell,1,k},\ldots,\tilde{X}_{\ell,n_T,k}]^T\in\mathbb{C}^{n_T\times 1}$. We assume that the interferers do not cooperate with the transmitters in the intended cell (Assumption~A2). We further invoke a random-coding argument and assume that the users within a cell draw their codebooks from the same distribution (Assumption~A3). This implies that the sequences $\{\tilde{\vect{X}}_{\ell,k},\,k\in\mathbb{Z}\}$, $\ell=0,1,\ldots$ are independent and each such sequence has the same distribution. Finally, the activity patterns $\{{B}_{\ell,u},\,u =1,\ldots n_T\}$, the additive noise  $\{\vect{Z}_k,\,k\in\mathbb{Z}\}$, and the fading coefficients $\{\vmat{H}_{\ell,k},\,k\in\mathbb{Z}\}$, $\ell=0,1,\ldots$ are assumed to be independent of each other. 

\subsection{Channel Capacity}

Let us denote a temporal sequence $S_{1}, S_{2}, \ldots, S_{n}$ by $S_{1}^n$.
We define the capacity of the channel model~\eqref{eqn:channel-model} as
\begin{equation}\label{Capacity-red0}
C(\mathsf{P}) \eqdef \lim_{n\rightarrow\infty}\frac{1}{n} \sup_{Q^n} I(\vect{X}_{0,1}^n;\vect{Y}_1^n).
\end{equation}
Here, the supremum is over all $n$-dimensional probability distributions $Q^n$ of $\tilde{\vect{X}}_{\ell,1}^n$, $\ell=0,1, 2,\ldots$,  satisfying the per-user average power constraint
\begin{eqnarray}\label{PowerC}
 \int{ \|\tilde{x}_{\ell,u,1}^n\|^2 dQ^n(\tilde{x}_{\ell,u,1}^n) }\leq n\mathsf{P},\  u=1,2,\ldots,n_T.
\end{eqnarray}
The logarithms used in this paper are natural logarithms. The capacity $C(\mathsf{P})$ has thus the dimension ``nats per channel use''. Intuitively, $C(\mathsf{P})$ characterizes the sum rate at which information can be transmitted within the intended cell. In order to obtain the rate per user, one would need to divide $C(\mathsf{P})$ by $n_T$.

We do not claim that there is a coding theorem associated with \eqref{Capacity-red0}, i.e., we do not claim that any rate below $C(\mathsf{P})$ is achievable in the sense that there exists an encoding and decoding scheme with this rate for which the decoding error probability tends to zero as $n$ tends to infinity. However, by Fano's inequality \cite[Sec.~7.9]{Cover}, any encoding and decoding scheme with a rate above $C(\mathsf{P})$ has a decoding error probability that is bounded away from zero as $n$ tends to infinity. Furthermore, it can be shown that any rate that follows by evaluating \eqref{Capacity-red0} for an i.i.d.\ distribution of $(\tilde{\vect{X}}_{\ell,1},\ldots,\tilde{\vect{X}}_{\ell,n})$ is achievable.

\section{Bounded Capacity}  \label{sec:bounded-capacity}

\subsection{Main Result}

We next present the main result of this paper: an upper bound on the channel capacity $C(\mathsf{P})$ that applies when the variances $\alpha_{\ell}$, $\ell=1,2,\ldots$ decay exponentially or more slowly. Since this upper bound does not depend on $\mathsf{P}$, it implies that, for such $\alpha_{\ell}$'s, the capacity is bounded in the signal-to-noise ratio ($\SNR$) of the system, defined as $\SNR \eqdef \mathsf{P} / \sigma^2$.

\begin{theorem}[Converse Bound]\label{thm:upper-bound}
Consider the channel model introduced in Section~\ref{sec:model}. Assume that, for some $0<\rho<1$,
\begin{equation}\label{eqn:alpha-ratio-bound}
    \frac{\alpha_{\ell+1}}{\alpha_{\ell}} \geq \rho, \quad \ell=1,2,\ldots 
\end{equation}
Then, for every $\mathsf{P}>0$ and $0 <\delta\leq 1$, the channel capacity is upper-bounded by
\begin{IEEEeqnarray}{lCl}
C(\mathsf{P})
& \leq & n_R\left((2-\delta)^{n_T}-1\right)\log\left(\rho^{-\frac{3}{2}}
\right)\nonumber\\
&&{}+\left(1-(1-\delta)^{n_T}\right)\left(\log\frac{\pi}{n_R\Gamma(n_R)} + n_R\log\frac{n_R}{e} + \frac{n_R}{2}\log (\eta_{\max})+ n_R\log(1+\eta_{\max})\right) \IEEEeqnarraynumspace \label{eqn:upper-bound}
\end{IEEEeqnarray}
where $\Gamma(\cdot)$ denotes the gamma function, and where we define
\begin{IEEEeqnarray}{lCl}
    \eta_{\max} \eqdef \max \left(\tfrac{1}{\alpha_1},\tfrac{1}{\rho}\right).\label{eqn:eta_max}
\end{IEEEeqnarray}
\end{theorem}
\begin{IEEEproof}
See Appendix~\ref{apx:proof-upper-bound}.
\end{IEEEproof}

Observe that the upper bound~\eqref{eqn:upper-bound} does not depend on the transmit power $\mathsf{P}$ and scales roughly linearly with the number of receive antennas at the base station $n_R$. Further observe that, for a fixed activity probability $\delta$, the bound grows exponentially with the number of users $n_T$ within the cell. However, this growth depends on $\delta$: when $\delta\to 0$, the bound is roughly proportional to $2^{n_T}$, whereas when $\delta\to 1$, the advantage of user cooperation within a cell vanishes.

Examining the proof of the upper bound in Appendix~\ref{apx:proof-upper-bound}, we find that the exponential growth in $n_T$  arises from the possibility of coordinated strategies among users within cells. More specifically, the definition of capacity \eqref{Capacity-red0} involves the maximization of a mutual information over all joint distributions of $\tilde{\mathbf{X}}_{\ell,1}^n$, $\ell=0,1,2,\ldots$, allowing for transmission strategies where $\tilde{X}_{\ell,u,k}$ are dependent and not necessarily identically distributed for $u=1,\ldots,n_T$. Suppose the set of active users in the interfering cells is different from the set of active users in the intended cell. In that case, this may be exploited to obtain a higher transmission rate, resulting in an exponential growth in $n_T$. However, this requires that we tailor the distribution of $\tilde{\mathbf{X}}_{\ell,1}^n$ to these sets of active users, which may not be realistic. Perhaps a more realistic setting arises by imposing that users behave independently of their index $u$, in which case the upper bound can be tightened, leading to a linear growth in $n_T$, as established in the next result.

We define the \emph{exchangeable capacity} as
\begin{equation}\label{eqn:exchangeable-capacity}
C_E(\mathsf{P}) \eqdef \lim_{n\rightarrow\infty}\frac{1}{n} \sup_{Q^n} I(\vect{X}_{0,1}^n;\vect{Y}_1^n),
\end{equation}
where the supremum is taken over all $n$-dimensional probability distributions $Q^n$ of 
$\tilde{\vect{X}}_{\ell,1}^n$, $\ell=0,1,2,\ldots$, that satisfy the per-user average power constraint \eqref{PowerC} and are \emph{exchangeable} in the sense that, for every permutation $\pi$ of $\{1,\ldots,n_T\}$,
\begin{equation}\label{eqn:perm-invariance}
(\tilde{X}_{\ell,1,1}^n,\ldots,\tilde{X}_{\ell,n_T,1}^n) \stackrel{d}{=} 
(\tilde{X}_{\ell,\pi(1),1}^n,\ldots,\tilde{X}_{\ell,\pi(n_T),1}^n).
\end{equation}
Note that the property of exchangeability is closely related to the concept of i.i.d.\ random variables: any sequence of random variables that are i.i.d.\ across users within a cell is indeed exchangeable.

\begin{corollary}[Exchangeable Capacity]\label{cor:upper-bound}
Assume that \eqref{eqn:alpha-ratio-bound} holds. Then, for every $\mathsf{P}>0$ and $0 < \delta \leq 1$, the exchangeable channel capacity is upper-bounded by
\begin{IEEEeqnarray}{lCl}
C_E(\mathsf{P})
& \leq & n_R n_T(1-\delta)\log\left(\rho^{-\frac{3}{2}}
\right)\nonumber\\
&&{}+\left(1-(1-\delta)^{n_T}\right)\left(\log\frac{\pi}{n_R\Gamma(n_R)} + n_R\log\frac{n_R}{e} + \frac{n_R}{2}\log (\eta_{\max})+ n_R\log(1+\eta_{\max})\right). \IEEEeqnarraynumspace \label{eqn:upper-bound-cor}
\end{IEEEeqnarray}
\end{corollary}

\begin{IEEEproof}
See Appendix~\ref{apx:proof-corollary-upper-bound}.
\end{IEEEproof}

\begin{figure}
    \centering
    \subfloat[Upper bounds as a function of the user activity probability $\delta$.]{
        \includegraphics[width=.48\textwidth]{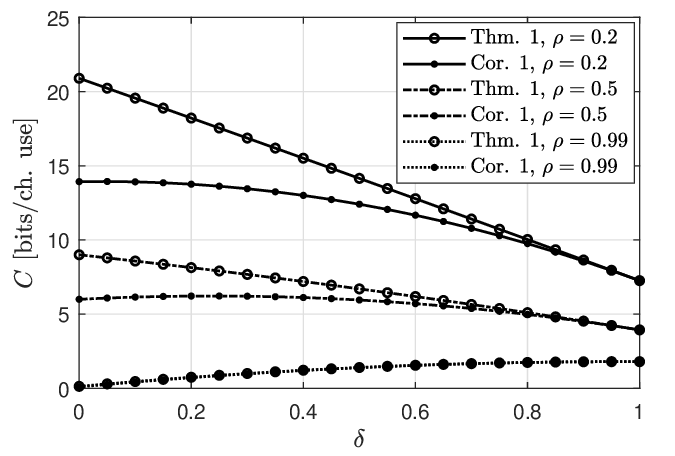}
        \label{fig:C_vs_delta}}
    \hfill
        \subfloat[Upper bounds as a function of the interference decay factor $\rho$.]{
        \includegraphics[width=.48\textwidth]{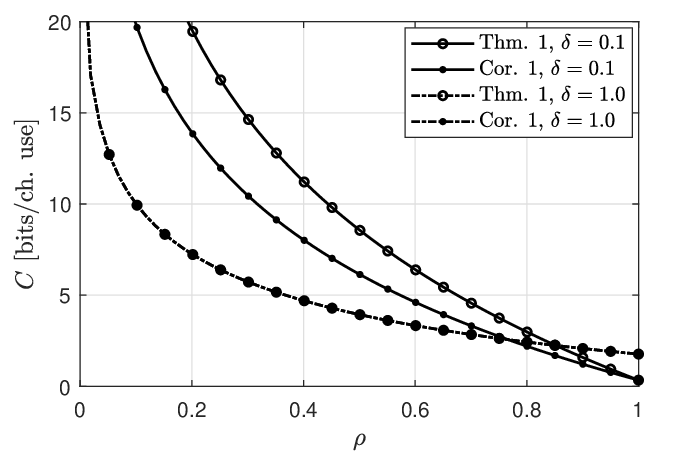}
        \label{fig:C_vs_rho}}
    \caption{Upper bounds on the channel capacity (Theorem~\ref{thm:upper-bound}) and the exchangeable channel capacity (Corollary~\ref{cor:upper-bound}) for $n_T=n_R=2$.}
    \label{fig:Th1}
\end{figure}

We observe that, in contrast to Theorem~\ref{thm:upper-bound}, the upper bound on the exchangeable capacity \eqref{eqn:upper-bound-cor} scales linearly with the number of transmitters $n_T$.
To illustrate the impact of user activity and the decay pattern of the variances of the fading coefficients on the capacity upper bounds, Figure~\ref{fig:Th1} presents the upper bounds on the channel capacity (Theorem~\ref{thm:upper-bound}) and on the exchangeable channel capacity (Corollary~\ref{cor:upper-bound}) for $n_T=n_R=2$. Specifically, Figure~\ref{fig:C_vs_delta} shows the upper bounds as a function of the user activity probability~$\delta$, while Figure~\ref{fig:C_vs_rho} depicts the upper bounds as a function of $\rho$, which is related to the decay of the variances of fading coefficients.

Observe that the impact of the user activity on the capacity upper bounds depends on $\rho$. In fact, the upper bounds suggest that, when $\rho$ is close to $1$, a small user activity is detrimental, whereas for small values of $\rho$, a small user activity is beneficial. Intuitively, a small user activity reduces the interference from other cells, but also reduces the time when users in the intended cell transmit information. Since the benefits of reduced interference are more pronounced when the variances of the fading coefficients decay faster, it follows that in this case, a reduced user activity is the most beneficial. 

Further observe that, when $\rho$ is close to $1$, the upper bounds are below 2.5 bits per channel use, irrespective of the user activity $\delta$. Intuitively, the assumption \eqref{eqn:alpha-ratio-bound} requires that the variances $\{\alpha_{\ell}\}$ decay at most exponentially. Indeed, letting $\alpha_0=1$, we have for every integer $L$ that
\begin{equation}\label{eq:exp_decay}
\frac{1}{L}\log \frac{1}{\alpha_L} = \frac{1}{L}\sum_{\ell=0}^{L-1}\log\frac{\alpha_{\ell}}{\alpha_{\ell+1}}.
\end{equation}
For any sequence $\{\alpha_{\ell}\}$ satisfying \eqref{eqn:alpha-ratio-bound}, the \emph{right-hand side (RHS)} of \eqref{eq:exp_decay} is upper-bounded by $\log(\eta_{\max})$, so the left-hand side of \eqref{eq:exp_decay} must be finite. This in turn implies that the sequence $\{\alpha_{\ell}\}$ decays exponentially or more slowly. In the limiting case as $\rho\to 1$, the decay of $\{\alpha_{\ell}\}$ becomes sub-exponential. As we shall argue in Section~\ref{sec:exponential-decay}, such a decay is consistent with most common propagation models. This suggests that, for $n_T=n_R=2$, the maximum rate achievable in noncoherent wireless networks does not exceed 2.5 bits per channel use.

Finally, observe that, as $\rho$ vanishes, our upper bounds grow to infinity. The case $\rho\to 0$ corresponds to the case where \eqref{eqn:alpha-ratio-bound} is not satisfied, i.e., where $\{\alpha_{\ell}\}$ decays faster than exponentially. In this scenario, Theorem~\ref{thm:upper-bound} does not apply and, consequently, the channel capacity may be unbounded. In Section~\ref{sec:bursty-signaling}, we show that this is the case when $\{\alpha_{\ell}\}$ decays faster than double-exponentially. 
Nevertheless, as we shall argue next, the assumption of an exponential or sub-exponential decay of $\{\alpha_{\ell}\}$ is reasonably mild and satisfied in most cases.

\subsection{Decay Patterns of Common Path Loss Models}  \label{sec:exponential-decay}

In practical wireless communication scenarios, most propagation models assume that the path loss increases polynomially with distance, capturing the attenuation of signal power as it propagates through space. Examples include the \emph{Free-Space Path Loss (FSPL)} model, applicable in idealized line-of-sight conditions; the two-ray model, which extends FSPL by incorporating ground-reflected signals; the single-slope path loss exponent model, which generalizes the polynomial decay to accommodate different propagation environments using an empirically determined exponent; and the Okumura-Hata model, based on extensive measurement data, among others \cite{Goldsmith2005,Jakborvornphan2020}.

Assume that the locations of interfering nodes can be modeled as a spatially random process over the plane. If the network cells follow a homogeneous Poisson point process with intensity $\eta$, the number of nodes in any region of area $A$ follows a Poisson distribution with mean $\eta A$, and the node locations are independent and uniformly distributed within that area. This statistical model accurately reflects scenarios where base stations or users are randomly deployed, such as in large-scale wireless networks or unplanned urban deployments.
The following proposition establishes a relation between the expected ordered distances $\d_\ell$ and the corresponding interfering cell index $\ell=1,2,\ldots$

\begin{proposition}[Ordered Distances]\label{lem:ordered-distances}
Let $x_0$ be a reference point in a $2$-dimensional plane where points are distributed according to a homogeneous Poisson point process $\Phi$ with intensity $\eta$. Denote by $D_\ell$ the distance from $x_0$ to the $\ell$-th closest point. Then, the expectation of this distance with respect to $\Phi$ satisfies
\begin{equation}
    \mathbb{E}_{\Phi}[D_\ell] = \sqrt{\frac{\ell}{\eta\pi}}.
\end{equation}
\end{proposition}
\begin{IEEEproof}
The number of points within a radius $r$ follows a Poisson distribution with mean $\eta \pi r^2$.  Therefore, the expected number of points $N(r)$ inside a ball of radius $r$ is given by
\begin{equation}\label{eqn:ordered-distances-1}
    \mathbb{E}_{\Phi}[N(r)] = \eta \pi r^2.
\end{equation}
Since $D_\ell$ is the distance to the $\ell$-th closest point, on average there should be $\ell$ points within a radius $r = \mathbb{E}_{\Phi}[D_\ell]$. Substituting this value in \eqref{eqn:ordered-distances-1} leads to the identity
\begin{equation}
    \ell = \eta \pi \mathbb{E}_{\Phi}[D_\ell]^2.
\end{equation}
Solving for $\mathbb{E}_{\Phi}[D_\ell]$ yields the desired result.
\end{IEEEproof}

By Proposition~\ref{lem:ordered-distances}, the distance from the $\ell$-th nearest interfering node to the intended node grows proportionally to $\sqrt\ell$. This scaling behavior arises from the expected spacing between points for a homogeneous Poisson point process with a certain intensity $\eta$. It establishes a direct relationship between the spatial distribution of interfering nodes and the decay rate of the associated fading coefficients. Since path loss typically follows an inverse polynomial dependence on distance, the ordered distances $d_\ell$ determine how fast the variances of the fading coefficients $\{\alpha_{\ell}\}$ decay with $\ell$. In particular, if the path-loss function $\text{PL}(d)$ follows a power-law behavior of the form  
\begin{equation}
  \text{PL}(d) \propto d^\beta,
\end{equation}
where $\beta$ is the path-loss exponent, then using Proposition~\ref{lem:ordered-distances}, we obtain that, for an average cell deployment,
\begin{equation}
\mathbb{E}_{\Phi}[\alpha_\ell] \approx \frac{1}{\text{PL}(\mathbb{E}_{\Phi}[D_\ell])} \propto \ell^{-\beta/2}.
\end{equation}
For instance, the free-space path loss model has $\beta=2$. Therefore, the expected fading variances $\{\alpha_{\ell}\}$ behave linearly with the cell-index $\ell$, $\mathbb{E}_{\Phi}[\alpha_\ell] \propto \ell^{-1}$.
We conclude that, for typical path loss models, the variances of the fading coefficients $\{\alpha_{\ell}\}$ decay slower than exponentially for an average cell deployment, and Theorem~\ref{thm:upper-bound} implies that the network has bounded capacity.

\begin{figure}
\centering
    \includegraphics[width=.45\textwidth]{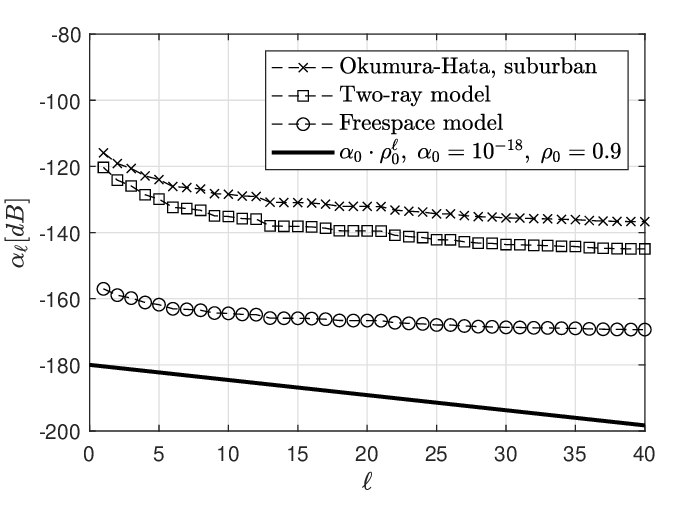}
\caption{Decay patterns with respect to the cell index $\ell$ under typical path-loss models for one realization of the Poisson point process $\Phi$ with intensity $\eta=3.2$ cells per square km.}\label{fig:fig-pathloss-models}
\end{figure}

Figure~\ref{fig:fig-pathloss-models} illustrates the fading variances (in dB) predicted by different path loss models for one realization of cells that are randomly deployed on a two-dimensional surface with a density of $3.2$ cells per square kilometer. The parameters for the Okumura-Hata model (suburban environment) are set as follows: frequency $f = 1.5$ GHz,  base-station height $h_b = 50$ meters, and mobile user height $h_m = 1.5$ meters. For the free-space and two-ray models, 
we use the same antenna heights, a link gain of $G = 10^{-4}$, and a transmission frequency of $f = 2.4$ GHz. For comparison, we further illustrate the fading variances $\alpha_{\ell} = 10^{-18}(0.9)^{\ell}$, which results in an exponential decay of $\{\alpha_{\ell}\}$ with $\eta_{\max} = 10^{18}$. We observe that, due to the polynomial behavior of the path loss in different models, the fading variances $\{\alpha_{\ell}\}$  decay more slowly than exponential.

\subsection{Fundamental Limits of Cooperation}
Lozano, Heath, and Andrews \cite{lozano_fundamental_2013} studied the fundamental limits of cooperation in wireless networks across a variety of scenarios. Among these, the case most relevant to our work is their analysis of the block-fading \emph{MIMO} channel model. In this model, a cell comprises $n_T$ transmitters and $n_R$ receivers, and the channel remains constant over $T$ symbol durations. When $n_T > T$, there are insufficient degrees of freedom within each coherence interval to estimate all channel coefficients independently, so this regime falls under the scope of noncoherent communication. They showed that, under these assumptions, the channel capacity is bounded in the $\SNR$.

In order to compare our results with those of \cite{lozano_fundamental_2013}, we focus on the specific case where $T = 1$. In this case, the channel model considered in \cite{lozano_fundamental_2013} has essentially the input-output relation
\begin{equation}\label{Lozano_ch}
\vect{Y}_k= \sqrt{\mathsf{P}}\vmat{H}_k\vect{X}_k + \vect{Z}_k
\end{equation}
where $\vect{X}_k \in \mathbb{C}^{n_T \times 1}$ corresponds to the vector of symbols transmitted at time $k$ by the users inside the intended cell, $\vmat{H}_k \in \mathbb{C}^{n_R \times n_T}$ denotes the matrix of fading coefficients, and $\vect{Z}_k \in \mathbb{C}^{n_T \times 1}$ denotes the additive noise vector. The sequence of noise vectors $\{\vect{Z}_k\}$ is i.i.d.\ with independent, zero-mean, unit-variance, circularly-symmetric, Gaussian entries. Similarly, $\{\mathbb{H}_k\}$ is a sequence of i.i.d.\ matrices with independent, zero-mean, circularly-symmetric Gaussian entries having variance $g_{r,u}>0$, $r=1,\ldots,n_R$, $u=1,\ldots,n_T$; cf.\ \cite[Eq. (4)]{lozano_fundamental_2013}. The vector of transmitted symbols $\vect{X}_k$ is assumed to be independent of $\mathsf{P}$ and to have unit-energy entries. As can be observed by comparing \eqref{Lozano_ch} with \eqref{eqn:channel-model}, in the channel model considered in \cite{lozano_fundamental_2013}, out-of-cluster interference is incorporated into the additive noise $\vect{Z}_k$. Furthermore, the channel input $\sqrt{\mathsf{P}}\vect{X}_k$ in \eqref{Lozano_ch} corresponds to $\vect{X}_k$ in \eqref{eqn:channel-model}. Lozano, Heath, and Andrews obtained the following upper bound on the information rate $I(\vect{X}_k;\vect{Y}_k)$.

\begin{proposition}[Lozano, Heath, and Andrews \cite{lozano_fundamental_2013}]\label{Lozano_Prop}
Consider the channel model \eqref{Lozano_ch} of a cellular system with $n_T > 1$ transmitting users. Define the diagonal matrices $\mat{G}_r=\textnormal{diag}(g_{r,1},\ldots,g_{r,n_T})$, $r=1,\ldots,n_R$.  Assume that the fading variances satisfy $g_{r,1}+\ldots+g_{r,n_T}=1$ for every $r=1,\ldots,n_R$ and that the vector of transmitted symbols $\vect{X}_k$ has non-zero entries almost surely. Then
\begin{IEEEeqnarray}{lCl}
    \varlimsup_{\mathsf{P}\to\infty} I(\vect{X}_k;\vect{Y}_k) \leq -\sum_{r=1}^{n_R} \E{\log ( \vect{X}_k^{\dagger}\mat{G}_r\vect{X}_k)}.\label{Lozano_Bound}
\end{IEEEeqnarray}
\end{proposition}
\begin{IEEEproof}
    See \cite[App. A]{lozano_fundamental_2013}.
\end{IEEEproof}

The bound \eqref{Lozano_Bound} has a similar flavor as our upper bound \eqref{eqn:upper-bound} in Theorem~\ref{thm:upper-bound}: both bounds imply that channel capacity (under some constraints on the channel inputs) is bounded in the transmit power. However, \eqref{Lozano_Bound} requires more stringent conditions on $\vect{X}_k$ than \eqref{eqn:upper-bound}. Among other things, the condition that $\vect{X}_k$ is independent of $\mathsf{P}$ is rather restrictive. In fact, it was shown by Lapidoth and Moser \cite[Th.~4.3]{lapidoth2003capacity} that the information rates achievable over \eqref{Lozano_ch} with inputs of the form $\sqrt{\mathsf{P}}\vect{X}_k$ (where $\vect{X}_k$ is independent of $\sqrt{P}$) are bounded in the transmit power. Thus, such inputs result in a bounded information rate even for a single-user channel. In other words, the boundedness of the information rate $I(\vect{X}_k;\vect{Y}_k)$ may be a consequence of the suboptimal input distribution rather than of the interference from other cells.
Furthermore, the condition that $\vect{X}_k$ has non-zero entries almost surely is not compatible with the random-access scenario considered in this paper, where users may be inactive with a positive probability. In contrast, our upper bound \eqref{eqn:upper-bound} in Theorem~\ref{thm:upper-bound} proves the boundedness of channel capacity even without imposing the aforementioned conditions on $\vect{X}_k$.

Intuitively, when users are inactive with a positive probability, the upper bound \eqref{Lozano_Bound} becomes infinite, and channel capacity is potentially unbounded. To see this, let us model the random user activity again by writing the channel inputs as
\begin{equation}
\label{bursty_input}
X_{u,k}=B_{u}\tilde{X}_{u,k}, \quad u=1,\ldots,n_T,
\end{equation}
where $\tilde{X}_{u,k}$ denotes the symbol transmitted by user $u$ at time $k$, and $B_u$ is a Bernoulli random variable that is $1$ with probability $\delta$ and $0$ otherwise. The random variable $B_u$ captures the activity of user $u$ and remains constant during the entire transmission. By following the same steps as in the proof of Proposition~\ref{Lozano_Prop}, we can derive the following upper bound analogous to \eqref{Lozano_Bound}.

\begin{proposition}\label{Lozano_bursty}
Consider the channel model \eqref{Lozano_ch} of a cellular system with $n_T > 1$ transmitting users. Let $\vect{X}_{k}=[X_{1,k},\ldots,X_{n_T,k}]^T$ be given by \eqref{bursty_input}, and assume that $\tilde{\vect{X}}_k = [\tilde{X}_{1,k},\ldots,\tilde{X}_{n_T,k}]^T$ is independent of $\mathsf{P}$ and its entries have unit energy and are non-zero almost surely. Further assume that the fading variances satisfy $g_{r,1}+\ldots+g_{r,n_T}=1$ for every $r=1,\ldots,n_R$. Then,
\end{proposition}
\begin{IEEEeqnarray}{lCl}
    I(\vect{X}_k;\vect{Y}_k) \leq n_R(1-\delta)^{n_T}\log\mathsf{P} + n_T H_b(\delta)+ n_R\log\left(1 + \frac{1}{\mathsf{P}}\right) - \left(1- (1-\delta)^{n_T}\right)\sum_{r=1}^{n_R} \E{\log(|\tilde{X}_{u_{\star}}|^2g_{r,u_{\star}})}, \label{eq:Lozano_bursty}
\end{IEEEeqnarray}
where $H_b(\cdot)$ denotes the binary entropy function and
\begin{equation}
u_{\star} \eqdef \arg\min_{u=1,\ldots,n_T} \E{\log(|\tilde{X}_{u}|^2g_{r,u})}.
\end{equation}
\begin{IEEEproof}
    See Appendix \ref{Proof_Lozano_bursty}.
\end{IEEEproof}

For $\delta<1$, the upper bound \eqref{eq:Lozano_bursty} grows logarithmically in $\mathsf{P}$ and fails to capture the boundedness of channel capacity in $\mathsf{P}$. However, in contrast to \eqref{eqn:upper-bound}, it does not depend on the decay rate of the variances of the fading coefficients $\{\alpha_{\ell}\}$. Furthermore, it suggests that, by artificially reducing the user activity, one may achieve an unbounded capacity when Theorem~\ref{thm:upper-bound} does not apply, i.e., when $\{\alpha_{\ell}\}$ decay faster than exponentially. We shall refer to this transmit strategy as \emph{bursty signaling} and show in the next section that it indeed achieves an unbounded information rate when $\{\alpha_{\ell}\}$ decay faster than double-exponentially.

\section{Bursty Signaling}  \label{sec:bursty-signaling}

In the previous section, we have seen that whenever the fading variances
$\{\alpha_{\ell}\}$ decay exponentially or sub-exponentially, noncoherent wireless networks have bounded capacity. In this section, we explore how to attain an unbounded coding rate if we relax this restriction.

Recall that the signal transmitted by user $u$ at time $k$ is modeled as
\begin{equation}
\label{eq:trans_signal}
X_{\ell,u,k} = B_{\ell,u} \tilde{X}_{\ell,u,k}, \quad \ell=0,1,\ldots
\end{equation}
where $B_{\ell,u}$ captures the activity of the user. We next follow a bursty signaling strategy to achieve an unbounded rate. To this end, consider an input distribution of the form
\begin{equation}
\label{eq:bursty_signaling}
\tilde{X}_{\ell,u,k} = \hat{B}_{\ell,u}\hat{X}_{\ell,u,k}, \quad \ell=0,1,\ldots
\end{equation}
where, for $u=1$, $\{\hat{B}_{\ell,u}\}$ are i.i.d.\ Bernoulli random variables with activation probability $\xi$ and $\{\hat{X}_{\ell,u,k}\}$ are i.i.d.\ circularly-symmetric random variables with $\log |\hat{X}_{\ell,u,k}|$ uniformly distributed over the interval $[0,\mathsf{P}]$; for $u=2,\ldots,n_T$, $\hat{B}_{\ell,u}$ and $\hat{X}_{\ell,u,k}$ are set to zero. Intuitively, the random variable $\hat{B}_{\ell,u}$ artificially reduces the user activity. This, in turn, reduces the interference from interfering cells to the intended cell. The distribution of $\hat{X}_{\ell,u,k}$ is the one that achieves the high-SNR asymptotic capacity of noncoherent single-user fading channels; cf.~\cite{lapidoth2003capacity}.

Let $L_{\mathsf{P}}\in\Naturals$ denote the number of interfering cells such that the total interference from the remaining cells is below the noise level, i.e., let $L_{\mathsf{P}}$ be such that
\begin{equation}\label{eqn:conditionL}
\sum_{\ell=L_{\mathsf{P}}+1}^{\infty}\alpha_{\ell}\mathsf{P} \leq \sigma^2.
\end{equation}
If there are only finitely many interfering cells, say $\mathsf{L}$, then $L_{\mathsf{P}} = \mathsf{L}$ independently of $\mathsf{P}$. Otherwise, the interference of the remaining cells increases with $\mathsf{P}$, so $L_{\mathsf{P}}$ must tend to infinity as $\mathsf{P}\to\infty$ to keep the interference below the noise level. The growth of $L_{\mathsf{P}}$ is determined by the decay rate of $\{\alpha_{\ell}\}$. For example, when $\{\alpha_{\ell}\}$ decays exponentially, the smallest $L_{\mathsf{P}}$ satisfying \eqref{eqn:conditionL}, denoted as $L_{\mathsf{P}}^{\star}$, grows logarithmically in $\mathsf{P}$. Indeed, when $\alpha_{\ell} = \rho^{\ell}$, $\ell=1,2,\ldots$ for some $0<\rho<1$, the left-hand side of \eqref{eqn:conditionL} becomes $\mathsf{P}\rho^{L_{\mathsf{P}}}/(1-\rho)$ and $L_{\mathsf{P}}^{\star}$ is given by
\begin{equation}
L^{\star}_{\mathsf{P}} = \left\lceil\frac{\log P/\sigma^2}{\log(1/\rho)}+ \frac{\log(1\!-\!\rho)}{\log(\rho)}-1\right\rceil.
\end{equation}
The behavior of $L_{\mathsf{P}}^{\star}$ for other decays of $\{\alpha_{\ell}\}$ can be obtained along similar lines. Of particular interest in this section is the double-exponential decay, which is considered in the following lemma.

\begin{lemma}[$L_{\mathsf{P}}$ for Double-Exponential Decay]\label{lem:double-exponential-decay}
For some $a \geq 1$, let the fading variances $\{\alpha_{\ell}\}$ satisfy
\begin{IEEEeqnarray}{lCl}\label{eqn:double-exponential-decay-alpha}
  \alpha_\ell &\leq& \frac{1}{\exp(\exp(\ell^{a}))}, \quad \ell=1,2,\ldots
\end{IEEEeqnarray}
Then, any integer $L_{\mathsf{P}}$ satisfying
\begin{IEEEeqnarray}{lCl}\label{eqn:double-exponential-decay-L}
  L_{\mathsf{P}} & > & \left(\log\log \left(\frac{\mathsf{P}}{\sigma^2} \frac{e^{-a}}{1-e^{-a}} \right)\right)^{\frac{1}{a}}
\end{IEEEeqnarray}
will also satisfy \eqref{eqn:conditionL}.
\end{lemma}
\begin{IEEEproof}
See Appendix~\ref{apx:double-exponential-decay}.
\end{IEEEproof}

Intuitively, introducing $L_{\mathsf{P}}$ allows us to consider an interference network with only $L_{\mathsf{P}}$ interfering cells by treating the remaining cells as noise. Following this strategy together with the bursty signaling scheme described in \eqref{eq:bursty_signaling}, we obtain the following lower bound on $C(\mathsf{P})$.

\begin{theorem}[Bursty-Signaling Lower Bound]\label{thm:lower-bound}
Consider the channel model from Section~\ref{sec:model}. Then, for every $\mathsf{P}>0$ and $0<\delta \leq 1$, the bursty signaling scheme \eqref{eq:bursty_signaling} with activation probability $\xi$ achieves the lower bound
\begin{IEEEeqnarray}{lCl}\label{eqn:lower-bound}
C(\mathsf{P})\geq \delta\xi(1-\delta\xi)^{L_{\mathsf{P}}}
\left(\log\log\mathsf{P} -\gamma - \log(e) -2\log\left(1+\sqrt{2}\sigma\right)\right), \IEEEeqnarraynumspace
\end{IEEEeqnarray}
where $L_\mathsf{P}$ is an integer satisfying \eqref{eqn:conditionL} and $\gamma$ denotes Euler's constant.
\end{theorem}
\begin{IEEEproof}
See Appendix~\ref{sec:proof-lower-bound}.
\end{IEEEproof}

As mentioned before, when there are only a finite number of interfering cells, $L_{\mathsf{P}}$ is bounded. In this case, the lower bound in Theorem~\ref{thm:lower-bound} implies that $C(\mathsf{P})$ is unbounded and grows at least double-logarithmically with $\mathsf{P}$. However, in general $L_{\mathsf{P}}$ grows with $\mathsf{P}$, in which case the term $(1-\delta\xi)^{n_TL_{\mathsf{P}}}$ vanishes. It then depends on the order of $L_{\mathsf{P}}$ whether the RHS of \eqref{eqn:lower-bound} is unbounded. For example, when $\{\alpha_{\ell}\}$ decays exponentially, $L_{\mathsf{P}}$ grows logarithmically with $\mathsf{P}$. It can then be shown that \eqref{eqn:lower-bound} vanishes as $\mathsf{P}\to\infty$, even if, for every $\mathsf{P}$, we optimize the lower bound over $\xi$. This is consistent with the upper bound presented in Theorem~\ref{thm:upper-bound}, which applies to this case and is bounded.

In contrast, when the sequence $\{\alpha_{\ell}\}$ decays faster than double-exponentially, the lower bound \eqref{eqn:lower-bound}, and hence also $C(\mathsf{P})$, are unbounded, as shown in the next corollary. To this end, we need to choose the parameter $\xi$, which specifies the extra burstiness introduced by the signal scheme \eqref{eq:bursty_signaling}, to be a decaying function of $\mathsf{P}$. More precisely, we choose
\begin{equation}
\xi = \bigl( \log \log \mathsf{P} \bigr)^{-\frac{1+\varepsilon}{a}},
\end{equation}
where $a>1$ is as in \eqref{eqn:double-exponential-decay-alpha} and determines the rate of $\{\alpha_{\ell}\}$, and $0<\varepsilon<a-1$ is an arbitrary parameter that determines the growth of the lower bound \eqref{eqn:lower-bound} with $\mathsf{P}$.

\begin{corollary}[Achievable rate for double-exponential decay]\label{cor:on-off-signaling-lower-bound}
Consider the channel model introduced in Section~\ref{sec:model}. Further assume that the fading variances $\{\alpha_{\ell}\}$ satisfy \eqref{eqn:double-exponential-decay-alpha} for some $a > 1$. Then, for every $\mathsf{P}>0$ and $0<\delta \leq 1$, the bursty signaling scheme \eqref{eq:bursty_signaling} achieves
\begin{IEEEeqnarray}{lCl}\label{eqn:on-off-signaling-lower-bound}
\varliminf_{\mathsf{P}\to\infty}
\frac{C(\mathsf{P})}{\delta\bigl(\log \log \mathsf{P} \bigr)^{1-\frac{1+\varepsilon}{a}}} & > &
0\IEEEeqnarraynumspace
\end{IEEEeqnarray}
for every $0<\varepsilon<a-1$.
\end{corollary}
\begin{IEEEproof}
See Appendix~\ref{apx:on-off-signaling-lower-bound}.
\end{IEEEproof}

Corollary~\ref{cor:on-off-signaling-lower-bound} implies that the capacity $C(\mathsf{P})$ is asymptotically lower-bounded by
\begin{equation}
\label{eqn:on-off-signaling-lower-bound-R}
C(\mathsf{P}) \gtrapprox \delta \bigl(\log \log \mathsf{P} \bigr)^{\eta}
\end{equation}
for every $0<\eta<(a-1)/a$. Clearly, this lower bound is unbounded, even though it grows only double-logarithmically with $\mathsf{P}$. In any case, the double-logarithmic growth is inherent to the memoryless noncoherent fading model and does not stem from the bursty signaling scheme \eqref{eq:bursty_signaling}. In fact, even in the absence of interfering cells, the capacity grows only double-logarithmically with $\mathsf{P}$; see \cite{lapidoth2003capacity}. We believe the approach of introducing signal burstiness to reduce inter-cell interference may be a promising transmission strategy at high SNR, where interference is the most harmful.

\section{Conclusions}
\label{sec:discussion}

We modeled a cellular network as a fading channel with an infinite number of interfering cells, where users from different cells do not cooperate but are constrained to use codebooks generated according to the same distribution. We considered a noncoherent scenario in which users and receivers know only the statistics of the fading coefficients, but not their realizations---an assumption that we believe is realistic when the number of users and cells is large. We further assumed that users access the network at random with a positive activity probability $\delta$. This model combines the network model of Lozano, Heath, and Andrews \cite{lozano_fundamental_2013} with the assumptions of random user activity and an infinite number of interferers, as considered in the massive random access literature. To avoid that the number of bits per user and channel use vanishes as the blocklength tends to infinity, we account for the path loss of interfering cells, ensuring that interference from distant cells diminishes. Overall, this model captures the fundamental features of both eMBB and mMTC scenarios.

For this channel model, we derived rigorous upper and lower bounds on the capacity, i.e., the largest rate at which information can be transmitted reliably. The behavior of these bounds depends critically on the decay of the fading variances $\{\alpha_{\ell}\}$ that model the path loss from interfering cells. Specifically, our bounds demonstrate that if the fading variances decay at an exponential rate or slower, then capacity is bounded in the transmit power. This is the case in typical scenarios where the path loss
grows polynomially with the distance and the density of the interfering cells is uniform. In contrast, if the fading variances decay faster than double-exponentially, then capacity becomes unbounded with transmit power, as it occurs when the number of interfering cells is finite. The case where the variances decay faster than exponentially but slower than double-exponentially remains open.

Our bounds further allowed us to investigate the effect of intermittent user activity. Intuitively, lower activity reduces interference, yet when the fading variances decay exponentially or slower, burstiness alone cannot overcome the boundedness of capacity. In such situations, carefully designed frequency reuse schemes may be needed to accelerate the effective decay of the fading variances. Nevertheless, user burstiness does influence the absolute value of our upper bound. For example, for $\rho=0.5$, $\delta=1$, and $n_T=n_R=2$, the bound in Theorem~\ref{thm:upper-bound} is $3.94$ bits per channel use, whereas for $\rho=0.5$, $\delta=0.5$, and $n_T=n_R=2$, it increases to $6.70$ bits per channel use. This suggests that bursty signaling strategies, which deliberately reduce user activity, can enlarge capacity. Indeed, by combining bursty signaling with \emph{treating interference as noise (TIN)}, we showed that when the fading variances decay faster than double-exponentially, the capacity is unbounded in the transmit power. 

These insights naturally extend to many-access and massive random access scenarios. The sublinear scaling result of \cite{ChenCG17}---which states that each user can transmit only a sublinear number of bits in blocklength $n$ as the number of users grows---relies crucially on assuming equal fading variances for all interferers, effectively placing them at comparable distances. Our analysis shows that alternative decay patterns of $\alpha_\ell$ fundamentally reshape the interference landscape, suggesting that the sublinear scaling observed in \cite{ChenCG17} does not directly apply to other spatial models and that, in specific scenarios, we could even achieve an unbounded capacity.

In essence, our results complement the work of Lozano, Heath, and Andrews~\cite{lozano_fundamental_2013}, who analyzed a different network model and showed that capacity is also bounded in transmit power under certain constraints on the input distributions. In particular, \cite{lozano_fundamental_2013} requires that the channel inputs are from a scale family---i.e., the inputs can be written as $\sqrt{P}\vect{X}_k$ for a vector $\vect{X}_k$ that is independent of $\mathsf{P}$---and the entries of $\vect{X}_k$ have unit energy and are non-zero almost surely. This excludes bursty signaling strategies, such as the ones we employed in the proof of Corollary~\ref{cor:on-off-signaling-lower-bound} to show unbounded capacity when the variances of the fading coefficients decay faster than exponentially. In fact, it was shown in \cite[Th.~4.3]{lapidoth2003capacity} that, for noncoherent point-to-point fading channels, the rate achievable by any scale family of input distributions is bounded in the transmit power. We therefore argue that the constraints on the channel inputs imposed in \cite{lozano_fundamental_2013} are rather restrictive and could potentially be the reason for a bounded capacity. Our results show that this is not the case, unless the variances of the fading coefficients decay faster than exponentially: our upper bound presented in Theorem~\ref{thm:upper-bound} does not require these constraints on the channel inputs and is still bounded in the transmit power. This confirms the existence of a saturation regime in interference-limited networks, where capacity reaches a ceiling that is independent of power.

\appendices

\section{Proof of Theorem~\ref{thm:upper-bound}}\label{apx:proof-upper-bound}

\subsection{Proof Outline}

The proof of Theorem~\ref{thm:upper-bound} consists of various parts, as we shall outline below:

\subsubsection{Introducing $\mathcal{J}$-interfering cells} We begin by defining $\mathcal{J}$ as the random set of active users in the intended cell, i.e.,
\begin{equation}
\mathcal{J} \triangleq \bigl\{u=1,\ldots,n_T\colon B_{0,u}=1\bigr\}.
\end{equation}
 To simplify the analysis, we focus on a subset of interfering cells, termed \textit{$\mathcal{J}$-interfering cells}, where the set of active users includes $\mathcal{J}$. That is, the $\ell$-th interfering cell is said to be a $\mathcal{J}$-interfering cell if $\mathcal{J} \subseteq \mathcal{G}_{\ell}$,
 where
\begin{equation}
\mathcal{G}_{\ell} \triangleq \bigl\{u=1,\ldots,n_T \colon B_{\ell,u}=1\bigr\}.
\end{equation}
We next introduce the indicator variable $A_\ell$, which is $1$ if the $\ell$-th cell is $\mathcal{J}$-interfering and zero otherwise, i.e.,
\begin{equation}\label{eqn:A_ell-def}
    A_{\ell} =
      \begin{cases}
        1,& \mathcal{J} \subseteq \mathcal{G}_{\ell},\\
        0,&\text{otherwise.}
      \end{cases}
\end{equation}
To derive an upper bound, we assume a ``genie'' provides side information about the interference from users not in $\mathcal{J}$ of every interfering cell, as well as from users in $\mathcal{J}$ of non-$\mathcal{J}$-interfering cells. The side information enables the receiver to remove this interference, leading to the initial upper bound
    \begin{IEEEeqnarray}{lCl}
        I(\vect{X}_{0,1}^n;\vect{Y}_1^n)  &\leq& H(\mathcal{J}) + I(\vect{X}_{0,1}^n;\vect{\hat{Y}}_1^n\,|\,{A}_1^L, \mathcal{J}),
    \end{IEEEeqnarray}
    where $L$ is an arbitrary integer that we will let go to infinity at the end of the proof, and $\vect{\hat{Y}}_1^n$ represents the received signal when in each cell either all users in $\mathcal{J}$ are active or none of them is active, i.e.,
    \begin{equation}\label{eqn:channel-model-hat}
\vect{\hat{Y}}_k= \vmat{H}_{0,k}\vect{X}_{0,k} + \sum_{\ell=1}^{\infty} A_{\ell} \vmat{H}_{\ell,\mathcal{J},k} \vect{X}_{\ell,\mathcal{J},k} + \vect{Z}_k.
\end{equation}
Here and throughout this appendix, we use the notation $\vmat{H}_{\ell,\mathcal{A},k}$ to denote a matrix with the columns $a \in \mathcal{A}$ of $\vmat{H}_{\ell,k}$ and $\vect{X}_{\ell,\mathcal{A},k}$ to denote the vector with the components $a \in \mathcal{A}$ of the vector $\vect{X}_{\ell,k}$.

\subsubsection{Decomposition into differential entropies} We express the conditional mutual information $I(\vect{X}_{0,1}^n;\vect{\hat{Y}}_1^n\,|\,{A}_1^L,\mathcal{J})$ using differential entropies and lower-bound $h(\vect{\hat{Y}}_1^n\,|\,\vect{X}_{0,1}^n,{A}_1^L,\mathcal{J})$ by conditioning on $\vmat{H}_{0,1}^n$. When the use that, conditioned on $(\vmat{H}_{0,1}^n,\vect{X}_{0,1}^n,{A}_1^L,\mathcal{J})$, $\vect{\hat{Y}}_1^n$ has the same distribution as
\begin{equation}
\bar{\vect{Y}}_{k}= \sum_{\ell=1}^{L} A_{\ell} \bar{\vmat{H}}_{\ell,\mathcal{J},k} \vect{X}_{\ell,\mathcal{J},k} + \sum_{\ell=L+1}^{\infty} \bar{A}_{\ell} \bar{\vmat{H}}_{\ell,\mathcal{J},k} \vect{X}_{\ell,\mathcal{J},k}+ \bar{\vect{Z}}_k, \label{eq:33}
\end{equation}
where $\{\bar{\vmat{H}}_{\ell,k}, k\in \mathbb{Z}\}$, $\{\bar{A}_{\ell}, \ell=1,2,\ldots\}$, and $\{\bar{\vect{Z}}_k, k\in\mathbb{Z}\}$ respectively have the same distributions as $\{\vmat{H}_{\ell,k}, k\in \mathbb{Z}\}$, $\{A_{\ell}, \ell=1,2,\ldots\}$, and $\{\vect{Z}_k, k\in\mathbb{Z}\}$ but are independent of them. It follows that
    \begin{IEEEeqnarray}{lCl}
        I(\vect{X}_{0,1}^n;\vect{\hat{Y}}_1^n\,|\,{A}_1^L,\mathcal{J})
        &\leq&  h(\vect{\hat{Y}}_1^n\,|\,{A}_1^L,\mathcal{J})-h(\bar{\vect{Y}}_1^n\,|\,{A}_1^L,\mathcal{J}). \IEEEeqnarraynumspace
    \end{IEEEeqnarray}

\subsubsection{Pairing activity patterns}
We next introduce a one-to-one mapping $f_L$ that pairs a given activity pattern $a_1^L$ in the first entropy with a ``shifted'' pattern $\tilde{a}_1^L = f_L(a_1^L)$ in the second entropy. This mapping ensures that both patterns have the same Hamming weight and therefore also the same probability. Furthermore, $\tilde{a}_1^L$ is a shifted version of the non-zero part of $a_1^L$ preceded by a $1$, which allows us to express $\bar{\vect{Y}}_1^n$ as a shifted version of $\vect{\hat{Y}}_1^n$.

\subsubsection{Applying a variational bound based on the information inequality}
We combine the two aligned entropy terms using the identity $h(U)-h(V)=h(U|V)-h(V|U)$. To upper bound the resulting expression, we apply the following variational bound that follows from the nonnegativity of relative entropy:

\begin{lemma}\label{LemmaGibbs}
Let $f$ and $g$ be two arbitrary \emph{probability density functions (pdf)s}. If $- \int f(x)\log f(x)\text{d}x$ is finite, then $- \int f(x)\log g(x)\text{d}x$ exists and
\end{lemma}
\begin{IEEEeqnarray}{lCl}\label{GibbsIne1}
-\int f(x)\log f(x)\text{d}x\leq-\int f(x)\log g(x)\text{d}x.
\end{IEEEeqnarray}
\begin{IEEEproof}
See \cite[Lemma 8.3.1]{AshBook}.
\end{IEEEproof}

A clever choice of $g(\cdot)$ allows us then to establish an upper bound on the difference of conditional differential entropies based on the expected power of the received signals.

\subsubsection{Average and limits} The final result is obtained by averaging the derived upper bound over all possible realizations of the activity patterns ${A}_1^L$. Subsequently, we take the limits as the number of interfering cells $L$ and the blocklength $n$ tend to infinity, thereby establishing an upper bound \eqref{eqn:upper-bound} in Theorem~\ref{thm:upper-bound}.

We now provide the detailed steps of each of the 5 parts described in this outline.

\subsection{Introducing $\mathcal{J}$-Interfering Cells}
 
Using the compact notation $S_{1}^n$ for the sequence $S_{1}, S_{2}, \ldots, S_{n}$, we can write the channel model \eqref{eqn:channel-model} as
\begin{equation}\label{eqn:channel-model-bis}
\vect{Y}_{1}^n= \vmat{H}_{0,1}^n\vect{X}_{0,1}^n+\sum_{\ell=1}^{\infty} \vmat{H}_{\ell,1}^n
\vect{X}_{\ell,1}^n + \vect{Z}_1^n.
\end{equation}

We begin by upper-bounding the mutual information as
\begin{IEEEeqnarray}{lCl}
    I(\vect{X}_{0,1}^n;\vect{Y}_1^n)&\stackrel{(a)}{\leq}& I(\mathcal{J}, \vect{X}_{0,1}^n;\vect{Y}_1^n)\nonumber\\
    &\stackrel{(b)}{=}& I(\mathcal{J};\vect{Y}_1^n)+I(\vect{X}_{0,1}^n;\vect{Y}_1^n\,|\,\mathcal{J})\nonumber\\
    &\stackrel{(c)}{\leq}& H(\mathcal{J})+I(\vect{X}_{0,1}^n;\vect{Y}_1^n\,|\,\mathcal{J})\label{eqn:IXY_J-1}
\end{IEEEeqnarray}
where $(a)$ follows by giving the set of active users in the intended cell $\mathcal{J}$ as extra information; $(b)$ follows by applying the chain rule of mutual information; and $(c)$ follows by upper-bounding the mutual information $I(\mathcal{J};\vect{Y}_1^n)$ by the entropy of the random variable $\mathcal{J}$. Note that the entropy $H(\mathcal{J})$ does not scale with the blocklength $n$.

We next upper-bound the conditional mutual information in \eqref{eqn:IXY_J-1} by giving extra information to the receiver. Specifically, for each $\mathcal{J}$-interfering cell, i.e., for every $\ell$ satisfying $A_{\ell}=1$, we provide the extra information $\bigl\{\vmat{H}_{\ell,\mathcal{J}^C,1}^n,\; \vect{X}_{\ell,\mathcal{J}^C,1}^n\bigr\}$. For the remaining cells, we provide the extra information $\bigl\{\vmat{H}_{\ell,1}^n,\; \vect{X}_{\ell,1}^n\bigr\}$.
This allows us to upper-bound the conditional mutual information as
\begin{IEEEeqnarray}{lCl}\label{eqn:IXY_J-2}
    I(\vect{X}_{0,1}^n;\vect{Y}_1^n \,|\,\mathcal{J})
    &\stackrel{(a)}{\leq}& I(\vect{X}_{0,1}^n;\vect{Y}_1^n , \bigl\{\vmat{H}_{\ell,\mathcal{J}^C,1}^n,\; \vect{X}_{\ell,\mathcal{J}^C,1}^n\bigr\}_{\ell: A_{\ell} = 1}, \bigl\{\vmat{H}_{\ell,1}^n,\; \vect{X}_{\ell,1}^n\bigr\}_{\ell: A_{\ell} = 0}\,|\,\mathcal{J})\nonumber\\
    &\stackrel{(b)}{=}& I(\vect{X}_{0,1}^n;  \bigl\{\vmat{H}_{\ell,\mathcal{J}^C,1}^n,\; \vect{X}_{\ell,\mathcal{J}^C,1}^n\bigr\}_{\ell: A_{\ell} = 1}, \bigl\{\vmat{H}_{\ell,1}^n,\; \vect{X}_{\ell,1}^n\bigr\}_{\ell: A_{\ell} = 0} \,|\,\mathcal{J}) \notag\\
    &&{}+\ I(\vect{X}_{0,1}^n;\vect{Y}_1^n\,|\, \mathcal{J}, \bigl\{\vmat{H}_{\ell,\mathcal{J}^C,1}^n,\; \vect{X}_{\ell,\mathcal{J}^C,1}^n\bigr\}_{\ell: A_{\ell} = 1}, \bigl\{\vmat{H}_{\ell,1}^n,\; \vect{X}_{\ell,1}^n\bigr\}_{\ell: A_{\ell} = 0})\nonumber\\
    &\stackrel{(c)}{=}& I(\vect{X}_{0,1}^n;\vect{\hat{Y}}_1^n\,|\,\mathcal{J}),
\end{IEEEeqnarray}
where $\vect{\hat{Y}}_1^n$ was defined in \eqref{eqn:channel-model-hat}. Step $(a)$ follows because giving extra information increases mutual information; step $(b)$ follows by applying the chain rule of mutual information; and step $(c)$ follows since the first mutual information is equal to zero given the independence of $\vect{X}_{0,1}^n$, $\vmat{H}_{\ell,1}^n$ and $\vect{X}_{\ell,1}^n$, $\ell>0$, and since the second mutual information is equal to $I(\vect{X}_{0,1}^n;\vect{\hat{Y}}_1^n\,|\,\mathcal{J})$.

The conditional mutual information in \eqref{eqn:IXY_J-2} can be further upper-bounded by giving the activity information ${A}_1^L$ of the first $L$ interfering cells, where $L$ is an arbitrary integer that we will let go to infinity at the end of the proof. This yields
\begin{IEEEeqnarray}{lCl}
  I(\vect{X}_{0,1}^n;\vect{\hat{Y}}_1^n\,|\,\mathcal{J}) 
   &\stackrel{}{\leq}&
      I(\vect{X}_{0,1}^n;\vect{\hat{Y}}_1^n,{A}_1^L \,|\, \mathcal{J} )\nonumber\\
   & \stackrel{(a)}{=}&
      I(\vect{X}_{0,1}^n;{A}_1^L\,|\,\mathcal{J})+ I(\vect{X}_{0,1}^n;\vect{\hat{Y}}_1^n\,|\,{A}_1^L,\mathcal{J}) \nonumber\\
   & \stackrel{(b)}{=}&
      I(\vect{X}_{0,1}^n;\vect{\hat{Y}}_1^n\,|\,{A}_1^L, \mathcal{J}),
      \IEEEeqnarraynumspace\label{eqn:IXY_J-3}
\end{IEEEeqnarray}
where $(a)$ follows from the chain rule of mutual information and $(b)$ follows because $A_1^L$, which is a function of $\vect{X}_{\ell,1}^n$, $\ell=1,\ldots,L$, is independent of $\vect{X}_{0,1}^n$.

\subsection{Decomposition into Differential Entropies}

We write
\begin{IEEEeqnarray}{lCl}\label{eqn:IXY_J-3bis}
    I(\vect{X}_{0,1}^n;\vect{\hat{Y}}_1^n\,|\,{A}_1^L, \mathcal{J})
      &=& \sum_{\jmath} \Pr\{\mathcal{J}=\jmath\} I(\vect{X}_{0,1}^n;\vect{\hat{Y}}_1^n\,|\,{A}_1^L,\mathcal{J}=\jmath),
\end{IEEEeqnarray}
where the sum is over all sets of active users $\jmath\subseteq\{1,\ldots,n_T\}$. When $\jmath$ is the empty set, $\vect{X}_{0,1}^n=\vect{0}_1^n$ and the mutual information $I(\vect{X}_{0,1}^n;\vect{\hat{Y}}_1^n\,|\,{A}_1^L,\mathcal{J}=\jmath)$ is zero. We next upper-bound this mutual information when $\jmath$ is not the empty set. As we shall see, this bound depends only on the number of active users $|\jmath|$, i.e., on the cardinality of $\jmath$. Specifically, we write the mutual information in terms of differential entropies and upper-bound the resulting expression as
\begin{IEEEeqnarray}{lCl}
I(\vect{X}_{0,1}^n;\vect{\hat{Y}}_1^n\,|\,{A}_1^L,\mathcal{J}=\jmath)
&\stackrel{}{=}&  h(\vect{\hat{Y}}_1^n\,|\,{A}_1^L,\mathcal{J}=\jmath)-h(\vect{\hat{Y}}_1^n\,|\,\vect{X}_{0,1}^n,{A}_1^L,\mathcal{J}=\jmath)\nonumber\\
&\stackrel{(a)}{\leq}&  h(\vect{\hat{Y}}_1^n\,|\,{A}_1^L,\mathcal{J}=\jmath)-h(\vect{\hat{Y}}_1^n\,|\,\vmat{H}_{0,1}^n, \vect{X}_{0,1}^n,{A}_1^L,\mathcal{J}=\jmath)\nonumber\\
&\stackrel{(b)}{=}&  h(\vect{\hat{Y}}_1^n\,|\,{A}_1^L,\mathcal{J}=\jmath)-h(\vect{\hat{Y}}_1^n-\vmat{H}_{0,1}^n\vect{X}_{0,1}^n\,|\,{A}_1^L,\mathcal{J}=\jmath),\label{eqn:IXY_J-4}
\end{IEEEeqnarray}
where $(a)$ follows because conditioning reduces entropy, and $(b)$ follows because, conditioned on $({A}_1^L,\mathcal{J})$, the sequence $\vect{\hat{Y}}_1^n-\vmat{H}_{0,1}^n\vect{X}_{0,1}^n$ is independent of $ (\vmat{H}_{0,1}^n,\vect{X}_{0,1}^n)$.

For a given realization ${A}_1^L={a}_1^L$, and setting $a_0\triangleq 1$ (because when $\jmath$ is non-empty, the intended cell is always active), we define
\begin{IEEEeqnarray}{lCl}
\vect{\hat{Y}}_k\!\left(a_1^L, \jmath\right) 
&\eqdef& \sum_{\ell=0}^{L}a_{\ell}\vmat{H}_{\ell,\jmath,k}\vect{X}_{\ell,\jmath,k}  +\sum_{\ell=L+1}^{\infty}A_{\ell}\vmat{H}_{\ell,\jmath,k}\vect{X}_{\ell,\jmath,k} +  \vect{Z}_k,\IEEEeqnarraynumspace
\label{eqn:Yk_a_hat}\\
\bar{\vect{Y}}_k\!\left(a_1^L, \jmath\right)
  &\eqdef& \sum_{\ell=1}^{L} a_{\ell} \bar{\vmat{H}}_{\ell,\jmath,k}\vect{X}_{\ell,\jmath,k} + \sum_{\ell=L+1}^{\infty} \bar{A}_{\ell} \bar{\vmat{H}}_{\ell,\jmath,k}\vect{X}_{\ell,\jmath,k} +\bar{\vect{Z}}_k. \IEEEeqnarraynumspace
\label{eqn:Yk_a_bar}
\end{IEEEeqnarray}
As in \eqref{eq:33}, $\{\bar{\vmat{H}}_{\ell,k}, k\in \mathbb{Z}\}$, $\{\bar{A}_{\ell}, \ell=1,2,\ldots\}$, and $\{\bar{\vect{Z}}_k, k\in\mathbb{Z}\}$ have the same distributions as $\{\vmat{H}_{\ell,k}, k\in \mathbb{Z}\}$, $\{A_{\ell}, \ell=1,2,\ldots\}$, and $\{\vect{Z}_k, k\in\mathbb{Z}\}$ but are independent of them.

We recall that the first sum in \eqref{eqn:Yk_a_hat} starts at $\ell=0$, while the 
the first sum in \eqref{eqn:Yk_a_bar} starts at $\ell=1$. The random variable $\vect{\hat{Y}}_k(a_1^L, \jmath)$ is precisely $\vect{\hat{Y}}_1^n$ conditioned on $A_1^L=a_1^L$ and $\mathcal{J}=\jmath$. Moreover, conditioned on $A_1^L=a_1^L$ and $\mathcal{J}=\jmath$, $\vect{\hat{Y}}_1^n-\vmat{H}_{0,1}^n \vect{X}_{0,1}^n$ has the same distribution as $\bar{\vect{Y}}_k\left(a_1^L, \jmath\right)$. 
Thus, the differential entropy terms in \eqref{eqn:IXY_J-4} can be written as
\begin{IEEEeqnarray}{rCl}
  h(\vect{\hat{Y}}_1^n\,|\,{A}_1^L,\mathcal{J}=\jmath) 
    & \stackrel{}{=} &  \sum_{a_1^L\in \set{B}_{L}} \Pr\{{A}_1^L=a_1^L\} \,h\bigl(\vect{\hat{Y}}_1^n(a_1^L, \jmath)\bigr),
    \IEEEeqnarraynumspace\label{eqn:IXY_J-4-H1}\\
  h(\vect{\hat{Y}}_1^n-\vmat{H}_{0,1}^n\vect{X}_{0,1}^n\,|\,{A}_1^L,\mathcal{J}=\jmath)\;
    &\stackrel{}{=}& \sum_{\tilde{a}_1^L\in \set{B}_{L}} \Pr\{A_1^L=\tilde{a}_1^L\} \,h\bigl(\bar{\vect{Y}}_1^n(\tilde{a}_1^L, \jmath)\bigr),\label{eqn:IXY_J-4-H2}
\end{IEEEeqnarray}
where ${\set{B}}_L \triangleq \{0,1\}^L$ denotes the set of all binary sequences of length $L$.

\subsection{Pairing Activity Patterns}

We will pair the sequences $a_1^L$ and $\tilde{a}_1^L$ in \eqref{eqn:IXY_J-4-H1} and \eqref{eqn:IXY_J-4-H2} according to a specific mapping $f_L$. To this end, we consider a partition of $\set{B}_L$ based on the position of the leading $1$ in each sequence. Specifically, for $m=1,\ldots,L+1$, we define
 \begin{equation}\label{eqn:BLm_def}
    {\set{B}}_L(m) \eqdef
    \begin{cases}
      \bigl\{a_1^L\in\set{B}_L: {a}_1^m=[{0}_1^{m-1},{1}]\bigr\}, & 1 \leq m \leq L,\\
      \{{0}_1^L\}, & m=L+1.
    \end{cases}
  \end{equation}
In words, ${\set{B}}_L(m)$ is the set of all binary sequences of length $L$ whose leading $1$ is in the $m$-th position. The sets $\set{B}_L(m)$, $m=1,\ldots,L+1$ are disjoint and define a partition of $\set{B}_L$.

\begin{proposition}\label{Prop_mapping}
There exists a one-to-one and onto mapping $f_{L}\colon\set{B}_{L}\rightarrow \set{B}_{L}$
such that, for every ${a}_1^L \in {\set{B}}_L$, the vector $\tilde{a}_1^L = f_{L}({a}_1^L)$ lies in $\set{B}_L(m)$ (for some $m$ determined by $a_1^L$) and satisfies $\|\tilde{a}_{1}^L\|_1 = \|a_{1}^{L}\|_1$ and  $\tilde{a}_1^L = [0_1^{m-1} , 1, a_1^{L-m}]$. Here $\|\cdot\|_p$, $p\geq 1$ denotes the $p$-norm. 
\end{proposition}
\begin{IEEEproof}
The proof is based on an explicit construction of the mapping $f_{L}$ that satisfies these properties.
In particular, Algorithm~\ref{alg1} generates a vector $\tilde{\vect{b}}$ such that the conditions in the proposition hold. The output of Algorithm \ref{alg1} satisfies $\tilde{\vect{b}}= [{0}_1^{m-1},1,{b}_1^{L-m}]$ by construction. The number of ones in the sequences $\vect{b}$ and $\tilde{\vect{b}}$ is the same, since the proposed algorithm just reorders the different positions of the original vector to generate the output vector. 
To prove that the mapping is one-to-one and onto, note that Algorithm \ref{alg2} recovers the original sequence $\vect{b}$ from the corresponding $\tilde{\vect{b}}$ for any $\vect{b}\in {\set{B}}_L$.
\end{IEEEproof}

\begin{algorithm}[!t]
 \KwData{Binary sequence $\vect{b}$ of length $L$ with Hamming weight $\|\vect{b}\|_1=W$}
 \KwResult{Binary sequence $\tilde{\vect{b}}= [{0}_1^{m-1},1,{b}_1^{L-m}]$ of length $L$ and  Hamming weight $\|\tilde{\vect{b}}\|_1=W$.}
 \eIf{$\vect{b}={0}_1^L$} 
{$\tilde{\vect{b}}=\vect{b}$}
 {
    $i\leftarrow$  take the position of the right-most 1 in sequence $\vect{b}$  \\
    $m \leftarrow L-i+1$ (length of ${b}_i^L$)\\
    $\tilde{\vect{b}}=[{0}_1^{m-1},1,{b}_1^{L-m}]$
 }
\caption{Mapping between binary sequences $\vect{b}$ and $\tilde{\vect{b}}$.\label{alg1}}
\end{algorithm}

\begin{algorithm}[!t]
 \KwData{Binary sequence $\tilde{\vect{b}}$ of length $L$ with Hamming weight $\|\tilde{\vect{b}}\|_1=W$}
 \KwResult{Binary sequence ${\vect{b}}= [{b}_{m+1}^{L},1,{0}_1^{m-1}]$ of length $L$ and  Hamming weight $\|{\vect{b}}\|_1=W$.}
 \eIf{$\tilde{\vect{b}}={0}_1^L$} 
{$\vect{b}=\tilde{\vect{b}}$}
 {
    $i\leftarrow$  take the position of the left-most 1 in sequence $\vect{b}$  \\
    $m \leftarrow i$: (length of ${b}_1^i$)\\
    ${\vect{b}}=[\tilde{{b}}_{m+1}^{L},1,{0}_1^{m-1}]$
 }
 \caption{Mapping between binary sequences $\tilde{\vect{b}}$ and $\vect{b}$.\label{alg2}}
\end{algorithm}

Using the mapping $f_L$ from Proposition~\ref{Prop_mapping}, we can reorganize the terms on the RHS of \eqref{eqn:IXY_J-4-H2} as
\begin{IEEEeqnarray}{lCl}
  h(\vect{\hat{Y}}_1^n-\vmat{H}_{0,1}^n\vect{X}_{0,1}^n\,|\,{A}_1^L,\mathcal{J}=\jmath)
    &\stackrel{}{=}& \sum_{a_1^L\in \set{B}_{L}} \Pr\{A_1^L=f_{L}({a}_1^L)\} \,h\bigl(\bar{\vect{Y}}_1^n(f_{L}({a}_1^L), \jmath)\bigr).
    \label{eqn:IXY_J-4-H2bis}
\end{IEEEeqnarray}
Substituting \eqref{eqn:IXY_J-4-H1} and \eqref{eqn:IXY_J-4-H2bis} back in \eqref{eqn:IXY_J-4}, we obtain
\begin{IEEEeqnarray}{lCl}
I(\vect{X}_{0,1}^n;\vect{\hat{Y}}_1^n\,|\,{A}_1^L,\mathcal{J} = \jmath)  &\leq& \sum_{a_1^L\in \set{B}_{L}}   \Pr\{{A}_1^L=a_1^L\} \left[  h\bigl(\vect{\hat{Y}}_1^n(a_1^L, \jmath)\bigr) - h\bigl(\bar{\vect{Y}}_1^n(f_{L}({a}_1^L), \jmath)\bigr)\right],
\IEEEeqnarraynumspace\label{eqn:IXY_J-5}
\end{IEEEeqnarray}
where we used that the random variables $A_k$, $k=1,\ldots,L$ are i.i.d., and therefore according to Proposition~\ref{Prop_mapping}, $\|f_{L}({a}_1^L)\|_1 = \|a_{1}^{L}\|_1$, so $\Pr\{A_1^L=f_{L}({a}_1^L)\} = \Pr\{{A}_1^L=a_1^L\}$.

We next recall that the sets $\set{B}_{L}(m)$, $m=1,\ldots,L+1$ partition the set $\set{B}_L$. We can thus rewrite \eqref{eqn:IXY_J-5} as
\begin{IEEEeqnarray}{lCl}
I(\vect{X}_{0,1}^n;\vect{\hat{Y}}_1^n\,|\,{A}_1^L,\mathcal{J}=\jmath) 
& \leq &
\sum_{m=1}^{L+1} \sum_{{a}_1^L: f_{L}({a}_1^L)\in \set{B}_{L}(m)}   \Pr\{{A}_1^L=a_1^L\} \left[  h\bigl(\vect{\hat{Y}}_1^n(a_1^L, \jmath)\bigr) - h\bigl(\bar{\vect{Y}}_1^n(f_{L}({a}_1^L), \jmath)\bigr)\right].
\label{eqn:IXY_J-5bis}
\end{IEEEeqnarray}
By Assumption A3, the distribution of the random variables $\vect{X}_{\ell,1}^n$ does not depend on $\ell=0,1,\ldots$
Therefore, for every $f_L(a_1^L) \in \set{B}_{L}(m)$,
the random variable  $\bar{\vect{Y}}_k(f_L(a_1^L), \jmath)$ has the same distribution as
\begin{IEEEeqnarray}{lCl}
\bar{\vect{Y}}_k(a_1^L,\jmath, m) &\eqdef & \sum_{\ell=0}^{L-m} a_{\ell}\bar{\vmat{H}}_{\ell+m,\jmath,k}\vect{X}_{\ell,\jmath,k} + \sum_{\ell=L-m+1}^{\infty} \bar{A}_{\ell}\bar{\vmat{H}}_{\ell+m,\jmath,k}\vect{X}_{\ell,\jmath,k} + \bar{\vect{Z}}_k.\IEEEeqnarraynumspace
\label{eqn:Yk_a_bar_m}
\end{IEEEeqnarray}
Comparing \eqref{eqn:Yk_a_hat} and \eqref{eqn:Yk_a_bar_m}, we observe that, as $L\to\infty$, these two random variables are roughly equivalent except for a shift from $\ell$ to  $\ell + m$ in the fading matrices. We use this property next to bound the differential entropies in \eqref{eqn:IXY_J-5bis}.

\subsection{Applying a Variational Bound Based on the Information Inequality}

We focus on the bracketed term in~\eqref{eqn:IXY_J-5bis}. Using the equivalence in distribution of $\bar{\vect{Y}}_k(f_L(a_1^L),\jmath)$ and $\bar{\vect{Y}}_k(a_1^L,\jmath, m)$ for $f_{L}({a}_1^L)\in \set{B}_{L}(m)$, and using the identity $h(U)-h(V)=h(U|V)-h(V|U)$, it follows that
\begin{IEEEeqnarray}{lCl}
{h\bigl(\vect{\hat{Y}}_1^n(a_1^L,\jmath)\bigr) - h\bigr(\bar{\vect{Y}}_1^n(f_L(a_1^L), \jmath)\bigr)} 
 & = & {h\bigl(\vect{\hat{Y}}_1^n(a_1^L, \jmath)\bigr) - h\bigr(\bar{\vect{Y}}_1^n({a_1^L, \jmath,m})\bigr)} \nonumber\\
 & = & h\bigl(\vect{\hat{Y}}_1^n(a_1^L,\jmath)\bigm|\bar{\vect{Y}}_1^n(a_1^L,\jmath,m)\bigr) - h\bigr(\bar{\vect{Y}}_1^n(a_1^L,\jmath,m)\bigm| \vect{\hat{Y}}_1^n(a_1^L,\jmath)\bigr)\nonumber\\
& \leq & \sum_{k=1}^n \bigl[h\bigl(\vect{\hat{Y}}_k(a_1^L,\jmath) \bigm| \bar{\vect{Y}}_{k}(a_1^L,\jmath,m)\bigr)\nonumber\\
&& \qquad {}-h\bigl(\bar{\vect{Y}}_k(a_1^L,\jmath,m)\bigm|\bar{\vect{Y}}_1^{{k-1}}(a_1^L,\jmath,m),\vect{\hat{Y}}_1^n(a_1^L,\jmath)\bigr)\bigr],
 \label{hYtildeY}
\end{IEEEeqnarray}
where the inequality follows from the chain rule and because conditioning reduces entropy.

To upper-bound \eqref{hYtildeY}, we next apply Lemma~\ref{LemmaGibbs} with a cleverly chosen $g(\cdot)$. Since we bound the conditional entropy given $\bar{\vect{Y}}_{k}(a_1^L,\jmath,m)$, we can choose a conditional pdf that depends on $\bar{\vect{Y}}_{k}(a_1^L,\jmath,m)$. Specifically, let $f_{\vect{\hat{Y}}_k|\bar{\vect{Y}}_k}$ denote the true conditional {pdf} of $\vect{\hat{Y}}_k(a_1^L,\jmath)$ given $\bar{\vect{Y}}_k(a_1^L,\jmath,m)$. Lemma~\ref{LemmaGibbs} allows us to upper-bound the conditional differential entropy of $\vect{\hat{Y}}_k(a_1^L,\jmath)$ given $\bar{\vect{Y}}_k(a_1^L,\jmath,m)$ by replacing $f_{\vect{\hat{Y}}_k|\bar{\vect{Y}}_k}$  by an auxiliary {pdf}  $g_{\vect{\hat{Y}}_k|\bar{\vect{Y}}_k}$. For every \mbox{$\bar{\vect{Y}}_k(a_1^L,\jmath,m) = \bar{\vect{y}}_k$}, we choose
\begin{eqnarray}\label{Cauchy_density_bis}
g_{\vect{\hat{Y}}_k|\bar{\vect{Y}}_k}(\vect{\hat{y}}_k|\bar{\vect{y}}_k)=\frac{n_R\sqrt{\beta} \Gamma(n_R)}{\pi^{n_R+1}\|\vect{\hat{y}}_k\|_2^{n_R}}\frac{1}{1+\beta\|\vect{\hat{y}}_k\|_2^{2n_R}}
\end{eqnarray}
with $\beta=1/\|\bar{\vect{y}}_k\|_2^{2n_R}$.
This is the density of a circularly-symmetric complex random variable whose magnitude is Cauchy distributed. A similar {pdf} has been used in \cite{koch_multipath_2010} to analyze frequency-dispersive fading channels.

Using \eqref{Cauchy_density_bis} in \eqref{GibbsIne1}, and since  $a^{n_R}+b^{n_R}\leq (a+b)^{n_R}$ for $a,b,\geq 0$, we obtain that
\begin{IEEEeqnarray}{lCl}
{h\bigl(\vect{\hat{Y}}_k(a_1^L,\jmath)\bigm|\bar{\vect{Y}}_k(a_1^L,\jmath,m)\bigr)}
& \leq & (n_R+1)\log \pi-\log n_R -\log \Gamma(n_R)\nonumber\\
&& {}+\frac{n_R}{2}\Big(\E{\log\|\vect{\hat{Y}}_k(a_1^L,\jmath)\|_2^{2}}-\E{\log\|\bar{\vect{Y}}_k(a_1^L,\jmath,m)\|_2^{2}}\Big)\nonumber\\
&& {}+n_R\E{\log\bigl(\|\vect{\hat{Y}}_k(a_1^L,\jmath)\|_2^{2}+\|\bar{\vect{Y}}_k(a_1^L,\jmath,m)\|_2^{2}\bigr)}.\label{+h1}
\end{IEEEeqnarray}

Next, we consider the second conditional entropy in \eqref{hYtildeY}. By conditioning on $\{\vect{X}_{\ell,\jmath,k}\}_{\ell=1}^{\infty}$ and $\{\bar{A}_{\ell}\}_{\ell=L-m+1}^{\infty}$, the random variable $\bar{\vect{Y}}_k(a_1^L,\jmath,m)$ is independent of $(\bar{\vect{Y}}_1^{{k-1}}(a_1^L,\jmath,m),\vect{\hat{Y}}_1^n(a_1^L,\jmath))$ and has a Gaussian distribution. Hence,
\begin{IEEEeqnarray}{lCl}
{h\bigl(\bar{\vect{Y}}_k(a_1^L,\jmath,m)\bigm|\bar{\vect{Y}}_1^{{k-1}}(a_1^L,\jmath,m),\vect{\hat{Y}}_1^n(a_1^L,\jmath)\bigr)}
 & \geq & h\bigl(\bar{\vect{Y}}_k(a_1^L,\jmath,m)\bigm|\{\vect{X}_{\ell,\jmath,k}\}_{\ell=1}^{\infty},\bar{A}_{L-m+1}^{\infty} \bigr)\nonumber\\
& = & n_R\log(\pi e)
+n_R \E{\log \bar{\mathsf{K}}(a_1^{L-m},\bar{A}_{L-m+1}^{\infty},\vect{X}_{0,\jmath,k}^{\infty})}, \label{-h2}
\end{IEEEeqnarray}
where
\begin{IEEEeqnarray}{lCl}\label{Cov}
\bar{\mathsf{K}}(a_1^{L-m},\bar{A}_{L-m+1}^{\infty},\vect{X}_{0,\jmath,k}^{\infty}) & \eqdef & 
\sum_{\ell=0}^{L-m}a_{\ell}\alpha_{\ell+m}\|\vect{X}_{\ell,\jmath,k}\|_2^2 +\!\sum_{\ell=L-m+1}^{\infty}\bar{A}_{\ell}\alpha_{\ell+m}\|\vect{X}_{\ell,\jmath,k}\|_2^2+\sigma^2.
\end{IEEEeqnarray} 
The inequality in \eqref{-h2} follows because conditioning reduces entropy.

Using \eqref{+h1} and \eqref{-h2},  we obtain that \eqref{hYtildeY} can be upper-bounded by
\begin{IEEEeqnarray}{lCl}
{h\bigl(\vect{\hat{Y}}_1^n(a_1^L,\jmath)\bigr) - h\bigr(\bar{\vect{Y}}_1^n(f_L(a_1^L), \jmath)\bigr)} 
&\leq &
\sum_{k=1}^{n}\biggl(\log \frac{\pi}{n_R\Gamma(n_R)}-n_R\log e \nonumber\\
&&\quad\quad\  {}+\frac{n_R}{2}\Bigl(\E{\log\|\vect{\hat{Y}}_k(a_1^L,\jmath)\|_2^{2}}-\E{\log\|\bar{\vect{Y}}_k(a_1^L,\jmath,m)\|_2^{2}}\Bigr)\nonumber\\
&&\quad\quad\ {}+n_R\E{\log(\|\vect{\hat{Y}}_k(a_1^L,\jmath)\|_2^{2}+\|\bar{\vect{Y}}_k(a_1^L,\jmath,m)\|_2^{2})}\nonumber\\
&&\quad\quad\ {}-n_R\Exp\Bigl[\log \bar{\mathsf{K}}(a_1^{L-m},\bar{A}_{L-m+1}^{\infty},\vect{X}_{0,\jmath,k}^{\infty})\Bigr]\biggr). \label{GUB_11}
\end{IEEEeqnarray}

To upper-bound the second line in \eqref{GUB_11}, we note that, conditioned on $\vect{X}_{\ell,\jmath,k}=\vect{x}_{\ell}$, $A_{\ell}=a_{\ell}$, and $\bar{A}_{\ell}=\bar{a}_{\ell}$, $\ell=0,1,\ldots$, both $\|\vect{\hat{Y}}_k(a_1^L,\jmath)\|_2^2$ and $\|\bar{\vect{Y}}_k(a_1^L,\jmath,m)\|_2^2$ have a chi-square distribution with $2n_R$ degrees of freedom.
Using \cite[Eq.~4.352]{Gradshteyn2007}, we thus obtain that
\begin{IEEEeqnarray}{lCl}
{\E{\log\|\vect{\hat{Y}}_k(a_1^L,\jmath)\|_2^2}-\E{\log\|\bar{\vect{Y}}_k(a_1^L,\jmath,m)\|_2^2}}
 & = & \E{\log\left(\frac{\mathsf{K}(a_1^L,A_{L+1}^{\infty},\vect{X}_{0,\jmath,k}^{\infty})}{\bar{\mathsf{K}}(a_1^{L-m},\bar{A}_{L-m+1}^{\infty},\vect{X}_{0,\jmath,k}^{\infty})}\right)},\label{EEdiff-1}
\end{IEEEeqnarray}
where
\begin{IEEEeqnarray}{lCl} 
{\mathsf{K}(a_1^L,A_{L+1}^{\infty},\vect{X}_{0,\jmath,k}^{\infty})} & \eqdef & \sum_{\ell=0}^{L}a_{\ell}\alpha_{\ell}\|\vect{X}_{\ell,\jmath,k}\|_2^2+\sum_{\ell=L+1}^{\infty}A_{\ell}\alpha_{\ell}\|\vect{X}_{\ell,\jmath,k}\|_2^2+\sigma^2.
\end{IEEEeqnarray}
The theorem's assumption \eqref{eqn:alpha-ratio-bound} implies that
\begin{equation}\label{eqn:alpha_ell_bound}
 \alpha_{\ell+m} \geq \frac{\rho^{m-1}}{\eta_{\max}} \alpha_{\ell}, \quad \ell=0,1,\ldots
\end{equation}
This allows us to lower-bound
\begin{IEEEeqnarray}{lCl}
\bar{\mathsf{K}}(a_1^{L-m},\bar{A}_{L-m+1}^{\infty},\vect{X}_{0,\jmath,k}^{\infty}) & \geq & \frac{\rho^{m-1}}{\eta_{\max}}\sum_{\ell=0}^{L-m}a_{\ell}\alpha_{\ell}\|\vect{X}_{\ell,\jmath,k}\|_2^2 +\frac{\rho^{m-1}}{\eta_{\max}}\sum_{\ell=L-m+1}^{\infty}\bar{A}_{\ell}\alpha_{\ell+m}\|\vect{X}_{\ell,\jmath,k}\|_2^2+\sigma^2 \nonumber\\
& \geq & \frac{\rho^{m-1}}{\eta_{\max}}\sum_{\ell=0}^{L-m}a_{\ell}\alpha_{\ell}\|\vect{X}_{\ell,\jmath,k}\|_2^2 +\frac{\rho^{m-1}}{\eta_{\max}}\sum_{\ell=L-m+1}^{\infty}\bar{A}_{\ell}\alpha_{\ell+m}\|\vect{X}_{\ell,\jmath,k}\|_2^2+\frac{\rho^{m-1}}{\eta_{\max}}\sigma^2,
\end{IEEEeqnarray}
where the second inequality follows because $\rho^{m-1}/\eta_{\max} \leq 1$, since $\eta_{\max} \geq 1/\rho$ by definition and $0<\rho<1$ by assumption. Substituting the resulting bound in \eqref{EEdiff-1}, we thus obtain that 
\begin{IEEEeqnarray}{lCl}
\IEEEeqnarraymulticol{3}{l}{{\E{\log\|\vect{Y}_k(a_1^L,\jmath)\|_2^2}
-\E{\log\|\bar{\vect{Y}}_k(a_1^L,\jmath,m)\|_2^2}}} \nonumber\\
\quad & \leq & \E{\log\left(\frac{\sum_{\ell=0}^{L}a_{\ell}\alpha_{\ell}\|\vect{X}_{\ell,\jmath,k}\|_2^2+\sum_{\ell=L+1}^{\infty}A_{\ell}\alpha_{\ell}\|\vect{X}_{\ell,\jmath,k}\|_2^2+\sigma^2}{\sum_{\ell=0}^{L-m}a_{\ell}\alpha_{\ell}\|\vect{X}_{\ell,\jmath,k}\|_2^2 +\sum_{\ell=L-m+1}^{\infty}\bar{A}_{\ell}\alpha_{\ell}\|\vect{X}_{\ell,\jmath,k}\|_2^2+\sigma^2}\right)} + \log \left( \frac{\eta_{\max}}{\rho^{m-1}} \right) \nonumber\\
& \stackrel{(a)}{\leq} & \E{\log\left(1+\frac{\sum_{\ell=L-m+1}^{L}a_{\ell}\alpha_{\ell}\|\vect{X}_{\ell,\jmath,k}\|_2^2+\sum_{\ell=L+1}^{\infty}A_{\ell}\alpha_{\ell}\|\vect{X}_{\ell,\jmath,k}\|_2^2}{\sigma^2}\right)} + \log \left( \frac{\eta_{\max}}{\rho^{m-1}} \right)\nonumber\\
 & \stackrel{(b)}{\leq} & \log \left( \frac{\eta_{\max}}{\rho^{m-1}} \right)+\zeta_{L,m,k},
\label{E+-}
\end{IEEEeqnarray}
where
\begin{IEEEeqnarray}{lCl}
\!\zeta_{L,m,k} & \eqdef & 
\sum_{\ell=L-m+1}^{L} \frac{a_{\ell}\alpha_{\ell}\Exp[  \|\vect{X}_{\ell,\jmath,k}\|_2^2]}{\sigma^2} +\sum_{\ell=L+1}^{\infty}\frac{\alpha_{\ell}\Exp[A_{\ell}\|\vect{X}_{\ell,\jmath,k}\|_2^2]}{\sigma^2}. \label{error1}
\end{IEEEeqnarray}
In \eqref{E+-}, $(a)$ follows by writing the fraction as the sum of two fractions, the first given by $\sum_{\ell=0}^{L-m} a_{\ell}\alpha_{\ell}\|\vect{X}_{\ell,\jmath,k}\|_2^2 + \sigma^2$ divided by the denominator and the second given by the remaining terms in the numerator divided by the denominator, and by lower-bounding the denominator of the first fraction by $\sum_{\ell=0}^{L-m}a_{\ell}\alpha_{\ell}\|\vect{X}_{\ell,\jmath,k}\|_2^2$ and the denominator of the second fraction by $\sigma^2$; $(b)$ follows by upper-bounding $\log(1+x) \leq x$, $x\geq 0$.

To upper-bound the third line in \eqref{GUB_11}, we use the law of iterated expectations and upper-bound the conditional expectation using Jensen's inequality:
\begin{IEEEeqnarray}{lCl}
\IEEEeqnarraymulticol{3}{l}{\E{\log\left(\|\vect{\hat{Y}}_k(a_1^L,\jmath)\|_2^{2}+\|\bar{\vect{Y}}_k(a_1^L,\jmath,m)\|_2^{2} \right)}}\nonumber\\
\quad &\leq & 
\Exp\Bigl[
\log\Bigl(
\Exp\Bigl[
\|\vect{\hat{Y}}_k(a_1^L,\jmath)\|_2^{2}
+\|\bar{\vect{Y}}_k(a_1^L,\jmath,m)\|_2^{2} \,\Big|\, \{\vect{X}_{\ell,\jmath,k}, \ell=0,1,\ldots\},A_{1}^{\infty},\bar{A}_{1}^{\infty} \Bigr]\Bigr)\Bigr]
\nonumber\\
&= & \Exp\Biggl[ \log\left(n_R\mathsf{K}(a_1^L,A_{L+1}^{\infty},\vect{X}_{0,\jmath,k}^{\infty}) + n_R\bar{\mathsf{K}}(a_1^{L-m},\bar{A}_{L-m+1}^{\infty},\vect{X}_{0,\jmath,k}^{\infty})\right)\Biggr].
\label{UB_sum_Ch}
\end{IEEEeqnarray}
We next use again \eqref{eqn:alpha_ell_bound} to upper-bound
\begin{IEEEeqnarray}{lCl}
\mathsf{K}(a_1^L,A_{L+1}^{\infty},\vect{X}_{0,\jmath,k}^{\infty}) & \leq & \frac{\eta_{\max}}{\rho^{m-1}}\sum_{\ell=0}^{L-m}a_{\ell}\alpha_{\ell+m}\|\vect{X}_{\ell,\jmath,k}\|_2^2+\sum_{\ell=L-m+1}^{L}a_{\ell}\alpha_{\ell}\|\vect{X}_{\ell,\jmath,k}\|_2^2 + \sum_{\ell=L+1}^{\infty}A_{\ell}\alpha_{\ell}\|\vect{X}_{\ell,\jmath,k}\|_2^2+\sigma^2 \nonumber\\
& \leq & \frac{\eta_{\max}}{\rho^{m-1}}\sum_{\ell=0}^{L-m}a_{\ell}\alpha_{\ell+m}\|\vect{X}_{\ell,\jmath,k}\|_2^2 + \frac{\eta_{\max}}{\rho^{m-1}} \sum_{\ell=L-m+1}^{\infty}\bar{A}_{\ell}\alpha_{\ell+m}\|\vect{X}_{\ell,\jmath,k}\|_2^2+\frac{\eta_{\max}}{\rho^{m-1}}\sigma^2 \nonumber\\
& & {} + \sum_{\ell=L-m+1}^{L}a_{\ell}\alpha_{\ell}\|\vect{X}_{\ell,\jmath,k}\|_2^2 + \sum_{\ell=L+1}^{\infty}A_{\ell}\alpha_{\ell}\|\vect{X}_{\ell,\jmath,k}\|_2^2 \nonumber\\
& = & \frac{\eta_{\max}}{\rho^{m-1}}\bar{\mathsf{K}}(a_1^{L-m},\bar{A}_{L-m+1}^{\infty},\vect{X}_{0,k}^{\infty}) + \sigma^2 \zeta_{L,m,k}
\end{IEEEeqnarray}
where the second inequality follows because the second sum on the second line is nonnegative and because $\eta_{\max}/\rho^{m-1}\leq 1$. It follows that \eqref{UB_sum_Ch} can be further upper-bounded as
\begin{IEEEeqnarray}{lCl}
{\E{\log\left(\|\vect{\hat{Y}}_k(a_1^L,\jmath)\|_2^2+\|\bar{\vect{Y}}_k(a_1^L,\jmath,m)\|_2^2\right)}}&\leq & \Exp\Biggl[\log\Biggl(n_R\left(1+\frac{\eta_{\max}}{\rho^{m-1}}\right)\bar{\mathsf{K}}(a_1^{L-m},\bar{A}_{L-m+1}^{\infty},\vect{X}_{0,k}^{\infty})+n_R\sigma^2\zeta_{L,m,k}\Biggr)\Biggr] \nonumber\\
& = & \log\left(1+\frac{\eta_{\max}}{\rho^{m-1}}\right) + \E{\log \bar{\mathsf{K}}(a_1^{L-m},\bar{A}_{L-m+1}^{\infty},\vect{X}_{0,k}^{\infty})} + \log n_R\nonumber\\
& & {} + \E{\log\left(1+\frac{\sigma^2 \zeta_{L,m,k}}{\left(1+\frac{\eta_{\max}}{\rho^{m-1}}\right)\bar{\mathsf{K}}(a_1^{L-m},\bar{A}_{L-m+1}^{\infty},\vect{X}_{0,k}^{\infty})}\right)} \nonumber\\
& \leq & \log\left(1+\frac{\eta_{\max}}{\rho^{m-1}}\right) + \E{\log \bar{\mathsf{K}}(a_1^{L-m},\bar{A}_{L-m+1}^{\infty},\vect{X}_{0,k}^{\infty})} + \log n_R + \zeta_{L,m,k}, \IEEEeqnarraynumspace\label{UB_sum_Ch_Res}
\end{IEEEeqnarray}
where the last step follows by bounding $\log(1+x)\leq x$, $x\geq 0$ and $(1+\eta_{\max}/\rho^{m-1})\bar{\mathsf{K}}(a_1^{L-m},\bar{A}_{L-m+1}^{\infty},\vect{X}_{0,k}^{\infty}) \geq \sigma^2$.

Combining \eqref{E+-} and \eqref{UB_sum_Ch_Res}  with \eqref{GUB_11}, we obtain
\begin{IEEEeqnarray}{lCl} 
{h\bigl(\vect{\hat{Y}}_1^n(a_1^L,\jmath)\bigr) - h\bigr(\bar{\vect{Y}}_1^n(f_L(a_1^L), \jmath)\bigr)} 
&\leq &
 n\Big[n_R (m-1)\log \left( \rho^{-\frac{3}{2}} \right) + \log\frac{\pi}{n_R\Gamma(n_R)} + n_R \log\frac{n_R}{e} \nonumber\\
& & {} \quad+ \frac{n_R}{2}\log (\eta_{\max})+n_R\log(\rho^{m-1}+\eta_{\max})\Big] + \frac{3}{2}n_R\sum_{k=1}^n \zeta_{L,m,k}\nonumber\\
& \leq &  n\Big[n_R (m-1)\log \left( \rho^{-\frac{3}{2}} \right) + \log\frac{\pi}{n_R\Gamma(n_R)} + n_R \log\frac{n_R}{e} \nonumber\\
& & {} \quad+ \frac{n_R}{2}\log (\eta_{\max})+n_R\log(1 + \eta_{\max}) + \frac{3}{2}n_R \bar{\zeta}_{L,m}\Big],\label{eqn:lemmaHH-ub}
\end{IEEEeqnarray}
where in the second step we upper-bounded $\rho^{m-1}$ in the fifth term by $1$, and we used the power constraint \eqref{PowerC} together with $\E {A_{\ell}} = p_{|\jmath|}$ to upper-bound
 \begin{IEEEeqnarray}{lCl}\label{eqn:zeta-ub}
 \frac{1}{n}\sum_{k=1}^n \zeta_{L,m,k} &\leq& \sum_{\ell=L-m+1}^{L}
 \frac{\alpha_{\ell}n_T\mathsf{P}}{\sigma^2}+\sum_{\ell=L+1}^{\infty}\frac{\alpha_{\ell}p_{|\jmath|} n_T\mathsf{P}}{\sigma^2}
 \triangleq \bar{\zeta}_{L,m}.
 \end{IEEEeqnarray}
 
Back to \eqref{eqn:IXY_J-5bis}, we note that the upper bound \eqref{eqn:lemmaHH-ub} depends on the position $m$ of the leading $1$ of $f_{L}({a}_1^L)$, but not on the specific activity pattern ${a}_1^L$. The probability that a length-$L$ binary sequence has the leading $1$ in the $m$-th position is equal to
\begin{IEEEeqnarray}{lCl}
\sum_{{a}_1^L: f_{L}({a}_1^L)\in \set{B}_{L}(m)} \Pr\{{A}_1^L=a_1^L\} &=& \Pr\{f_L(A_1^L)\in \set{B}_L(m)\} \nonumber\\
&=& p_{|\jmath|}(1-p_{|\jmath|})^{m-1}, \IEEEeqnarraynumspace\label{eqn:leading-one-prob}
\end{IEEEeqnarray}
since $A_{\ell}$, $\ell=1,\ldots,L$ follow an independent Bernoulli distribution with parameter $p_{|\jmath|}$, and since $f_L(A_1^L)$ has the same number of ones as $A_1^L$. Therefore, using \eqref{eqn:lemmaHH-ub}--\eqref{eqn:leading-one-prob} in \eqref{eqn:IXY_J-5bis}, we obtain that
\begin{IEEEeqnarray}{lCl}
\IEEEeqnarraymulticol{3}{l}{\frac{1}{n}I(\vect{X}_{0,1}^n;\vect{\hat{Y}}_1^n\,|\,{A}_1^L,\mathcal{J}=\jmath)} \nonumber\\
\quad & \leq & \sum_{m=1}^{L+1} p_{|\jmath|}(1-p_{|\jmath|})^{m-1} \Big[n_R (m-1)\log \left( \rho^{-\frac{3}{2}} \right) + \log\frac{\pi}{n_R\Gamma(n_R)}  +n_R \log \frac{n_R}{e} + \frac{n_R}{2}\log (\eta_{\max})+n_R\log(1+\eta_{\max})\Big] \nonumber\\
& & {} + \frac{3}{2} n_R \sum_{m=1}^{L+1} p_{|\jmath|}(1-p_{|\jmath|})^{m-1} \bar{\zeta}_{L,m}.\label{eqn:IXY_J-6}
\end{IEEEeqnarray}

\subsection{Average and Limits}
\label{apx:sub_avg_lim}
Using that
\begin{equation}
\label{eq:Bin_Sum}
 \sum_{m=1}^{\infty} p_{|\jmath|}(1-p_{|\jmath|})^{m-1}=1
\end{equation}
and \cite[Eq. 0.231]{Gradshteyn2007}
\begin{equation}
 \sum_{m=1}^{\infty} p_{|\jmath|}(1-p_{|\jmath|})^{m-1} (m-1) =\frac{1-p_{|\jmath|}}{p_{|\jmath|}}
\end{equation}
the first sum on the RHS of \eqref{eqn:IXY_J-6} converges to
\begin{IEEEeqnarray}{lCl}\label{eqn:IXY_J-7}
\Upsilon_{|\jmath|} \eqdef n_R \tfrac{1-p_{|\jmath|}}{p_{|\jmath|}}\log \left(\rho^{-\frac{3}{2}}\right)+\log\frac{\pi}{n_R\Gamma(n_R)}+n_R \log \frac{n_R}{e}+\frac{n_R}{2}\log \eta_{\max}+n_R\log\left(1+\eta_{\max}\right)
\end{IEEEeqnarray}
as $L\to\infty$.

We next show that the second sum on the RHS of~\eqref{eqn:IXY_J-6} vanishes as $L$ tends to infinity. Indeed, after some algebraic manipulations, we obtain that
\begin{IEEEeqnarray}{lCl}
\sum_{m=1}^{L+1} p_{|\jmath|}(1-p_{|\jmath|})^{m-1} \bar{\zeta}_{L,m} & = & \frac{n_T\mathsf{P}}{\sigma^2}\sum_{m=1}^{L+1}p_{|\jmath|}(1-p_{|\jmath|})^{m-1} \sum_{\ell=L-m+1}^L\alpha_{\ell} + \frac{n_T\mathsf{P}}{\sigma^2}\sum_{m=1}^{L+1}p^2_{|\jmath|}(1-p_{|\jmath|})^{m-1} \sum_{\ell=L+1}^{\infty}\alpha_{\ell} \nonumber\\
& \stackrel{(a)}{=} &  \frac{n_T\mathsf{P}}{\sigma^2}\sum_{\ell=0}^L \alpha_{\ell} \sum_{m=L-\ell+1}^{L+1} p_{|\jmath|}(1-p_{|\jmath|})^{m-1} + \frac{n_T\mathsf{P}}{\sigma^2}p_{|\jmath|}\sum_{\ell=L+1}^{\infty}\alpha_{\ell} \nonumber\\
& \stackrel{(b)}{\leq} & \frac{n_T\mathsf{P}}{\sigma^2}\sum_{\ell=0}^L \alpha_{\ell} (1-p_{|\jmath|})^{L-\ell} + \frac{n_T\mathsf{P}}{\sigma^2}p_{|\jmath|}\sum_{\ell=L+1}^{\infty}\alpha_{\ell},
\label{errorTerm1}
\end{IEEEeqnarray}
where $(a)$ follows by changing the order of the sums in the first term, and by evaluating the sum over $m$ in the second term using \eqref{eq:Bin_Sum}; to obtain $(b)$, we upper-bound the first term by letting the sum over $m$ run up to infinity, and by then evaluating the sum of a geometric series.

The second term on the RHS of \eqref{errorTerm1} vanishes as $L\to\infty$ since $\sum_{\ell=0}^{\infty}\alpha_{\ell} < \infty$ by assumption. To bound the first term on the RHS of \eqref{errorTerm1}, we fix an arbitrary $\nu$ and bound the sum as
\begin{IEEEeqnarray}{lCl}
\sum_{\ell=0}^L \alpha_{\ell} (1-p_{|\jmath|})^{L-\ell} & = & \sum_{\ell=0}^{\nu} \alpha_{\ell} (1-p_{|\jmath|})^{L-\ell} + \sum_{\ell=\nu+1}^{L} \alpha_{\ell} (1-p_{|\jmath|})^{L-\ell} \nonumber\\
& \leq & (1-p_{|\jmath|})^{L-\nu}\sum_{\ell=0}^{\infty} \alpha_{\ell} + \sum_{\ell=\nu+1}^{\infty}\alpha_{\ell}, \label{eq:zeta_nu_bound}
\end{IEEEeqnarray}
where we upper-bound the first sum by bounding $(1-p_{|\jmath|})^{L-\ell}\leq (1-p_{|\jmath|})^{L-\nu}$, the second sum by bounding $(1-p_{|\jmath|})^{L-\ell} \leq 1$, and both sums by changing the upper limit of the sum over $\ell$ to $\infty$. The first sum on the RHS of \eqref{eq:zeta_nu_bound} vanishes as $L\to\infty$, whereas the second sum vanishes as $\nu\to\infty$.

Combining \eqref{eq:zeta_nu_bound} with \eqref{errorTerm1}, we thus obtain that the first term on the RHS of \eqref{errorTerm1} vanishes as we first let $L\to\infty$ and then $\nu\to\infty$. As a result, the RHS of \eqref{eqn:IXY_J-6} tends to \eqref{eqn:IXY_J-7} as $L\to\infty$. Observe that \eqref{eqn:IXY_J-7} only depends on the set of active users in the intended cell $\jmath$ via its cardinality $|\jmath|$. We next use that, for $j \triangleq |\jmath|$, the activity pattern in the intended cell satisfies
\begin{IEEEeqnarray}{rCl}\label{eqn:PrJ-intended}
\Pr\{\mathcal{J}=\jmath\} &=& \delta^j(1-\delta)^{n_T-j}
\end{IEEEeqnarray}
and that there are $\binom{n_T}{j}$ of distinct sets $\jmath$ with cardinality $j$. Also, the probability of a cell being $\mathcal{J}$-interfering is 
\begin{IEEEeqnarray}{rCl}
  p_{|\jmath|} &=& \delta^j.\label{eqn:pjmath-J-interfering}
\end{IEEEeqnarray}
Then, using \eqref{eqn:PrJ-intended} and \eqref{eqn:pjmath-J-interfering},
we obtain from  \eqref{eqn:IXY_J-6}, \eqref{eqn:IXY_J-7}, and \eqref{eqn:IXY_J-3bis} that
\begin{IEEEeqnarray}{lCl}
\varlimsup_{L\to\infty}\frac{1}{n} I(\vect{X}_{0,1}^n;\vect{\hat{Y}}_1^n\,|\,{A}_1^L, \mathcal{J})
 &\leq &\sum_{\jmath}\Pr\{\mathcal{J}=\jmath\} \Upsilon_{|\jmath|}\nonumber\\
&= &\sum_{j=1}^{n_T} \binom{n_T}{j} \delta^j(1-\delta)^{n_T-j} n_R\frac{1-\delta^j}{\delta^j}\log \left( \rho^{-\frac{3}{2}} \right) \nonumber\\
& & {} + \left(1-(1-\delta)^{n_T}\right)\left(\log\frac{\pi}{n_R\Gamma(n_R)} + n_R\log\frac{n_R}{e}+\frac{n_R}{2}\log (\eta_{\max})+ n_R\log(1+\eta_{\max})\right) \label{MI_almost_fin}\IEEEeqnarraynumspace
\end{IEEEeqnarray}
Evaluating the first term on the RHS of \eqref{MI_almost_fin} using the binomial theorem, and combining \eqref{MI_almost_fin} with \eqref{eqn:IXY_J-1}--\eqref{eqn:IXY_J-3}, we obtain
\begin{IEEEeqnarray}{lCl}
\frac{1}{n}{I(\vect{X}_{0,1}^n;\vect{Y}_1^n)}&\leq& \frac{1}{n} H(\mathcal{J}) + n_R\left((2-\delta)^{n_T}-1\right)\log\left(\rho^{-\frac{3}{2}}
\right)\nonumber\\
&&{}+\left(1-(1-\delta)^{n_T}\right)\left(\log\frac{\pi}{n_R\Gamma(n_R)} + n_R\log\frac{n_R}{e}+\frac{n_R}{2}\log (\eta_{\max})+ n_R\log(1+\eta_{\max})\right)
 \label{MI_Fin}
\end{IEEEeqnarray}
from which the desired bound \eqref{eqn:upper-bound} follows by noting that, as $n\to\infty$, the term $\frac{1}{n} H(\mathcal{J})$ vanishes.

\section{Proof of Corollary~\ref{cor:upper-bound}}\label{apx:proof-corollary-upper-bound}

The proof of Corollary~\ref{cor:upper-bound} closely follows the proof of Theorem~\ref{thm:upper-bound} in Appendix~\ref{apx:proof-upper-bound}, but relies on a slightly different definition of interfering cells. 
Specifically, in the proof of Corollary~\ref{cor:upper-bound}, we say that a cell~$\ell$ is \emph{$J$-interfering} if 
\begin{equation}
  |\mathcal{J}| \leq |\mathcal{G}_{\ell}|,
\end{equation}
that is, if the number of active users in $\mathcal{G}_{\ell}$ is at least the number of active users in the intended cell $\mathcal{J}$.  We adopt the notation $J$-interfering (instead of $\mathcal{J}$-interfering) to emphasize that the definition depends only on the cardinality $J = |\mathcal{J}|$, rather than on the specific set $\mathcal{J}$.
In words, instead of restricting the analysis to cells whose active-user set $\mathcal{G}_{\ell}$ contains the entire set $\mathcal{J}$, we consider all interfering cells with at least $|\mathcal{J}|$ active users.

As in the proof of Theorem~\ref{thm:upper-bound}, we define the corresponding indicator variable $A_\ell$, which is $1$ if the $\ell$-th cell is $J$-interfering and zero otherwise, i.e.,
\begin{equation}\label{eqn:A_ell-def2}
    A_{\ell} =
      \begin{cases}
        1,& |\mathcal{J}| \leq |\mathcal{G}_{\ell}|,\\
        0,&\text{otherwise.}
      \end{cases}
\end{equation}
For the $J$-interfering cells, we apply a permutation on the user indices $u=1,\ldots,n_T$ on both $X_{\ell,u,k}$ and $H_{\ell,u,k}$ so that the active indices in cell $\ell$ are aligned with the active users in the intended cell $\mathcal{J}$. This permutation does not affect the mutual information $I(\vect{X}_{0,1}^n;\vect{Y}_1^n\,|\,\mathcal{J})$, since the transmitted symbols $(\tilde{X}_{\ell,1,1}^n,\ldots,\tilde{X}_{\ell,n_T,1}^n)$ are exchangeable by the corollary's assumption, and since $(H_{\ell,1,k},\ldots,H_{\ell,1,k})$ and $(B_{\ell,1},\ldots,B_{\ell,n_T})$ are i.i.d.\ for all $\ell = 0,1,\ldots$ according to the channel model.

After applying this permutation, all the steps in Appendix~\ref{apx:proof-upper-bound} apply for the new $J$-interfering cells until the final average of the bound in Section~\ref{apx:sub_avg_lim}. Indeed, let $\jmath$ be the set of active users in the intended cell and let $j=|\jmath|$. Then, the probability of the $\ell$-th cell being $J$-interfering is
\begin{equation}\label{eqn:Jinterfering-probability-cor}
    p_{|\jmath|} = \Pr[G \ge j] = \displaystyle \sum_{k=j}^{n_T} \binom{n}{k}\,\delta^k(1-\delta)^{n-k},
\end{equation}
where $G \sim \mathrm{Bin}(n_T,\delta)$ follows a binomial distribution with $n_T$ trials and activation probability $0<\delta<1$. Using \eqref{eqn:Jinterfering-probability-cor} in \eqref{errorTerm1}, instead of $p_{|\jmath|} = \delta^{j}$, the steps \eqref{errorTerm1}--\eqref{MI_almost_fin} then yield that
\begin{IEEEeqnarray}{lCl}
\varlimsup_{L\to\infty}\frac{1}{n} I(\vect{X}_{0,1}^n;\vect{\hat{Y}}_1^n\,|\,{A}_1^L, \mathcal{J})
 &\leq & n_R \sum_{j=1}^{n_T} \Pr[G = j] \frac{1-\Pr[G \ge j]}{\Pr[G \ge j]}\log \left( \rho^{-\frac{3}{2}} \right) \nonumber\\
& & {} + \left(1-(1-\delta)^{n_T}\right)\left(\log\frac{\pi}{n_R\Gamma(n_R)} + n_R\log\frac{n_R}{e}+\frac{n_R}{2}\log (\eta_{\max})+ n_R\log(1+\eta_{\max})\right).\IEEEeqnarraynumspace\label{eqn:MI_almost_fin_cor}
\end{IEEEeqnarray}
Therefore, to prove \eqref{eqn:upper-bound-cor} in Corollary~\ref{cor:upper-bound}, it only remains to show that $\sum_{j=1}^{n_T} \Pr[G = j] \frac{1-\Pr[G \ge j]}{\Pr[G \ge j]}$ is upper bounded by $n_T (1-\delta)$.
To this end, note that 
\begin{IEEEeqnarray}{lCl}
\sum_{j=1}^{n_T} \Pr[G = j] \frac{1-\Pr[G \ge j]}{\Pr[G \ge j]} 
  &\stackrel{(a)}{\leq}& \sum_{j=1}^{n_T} \Pr[G = j] \frac{1-\Pr[G \ge j]}{\Pr[G = j]}
  \nonumber\\
  &\stackrel{(b)}{=} & \sum_{j=1}^{n_T} \Pr[G < j]
  \nonumber\\
  &\stackrel{(c)}{=} & \sum_{k=0}^{n_T} (n_T-k)\,\Pr[G=k]
  \nonumber\\
  &\stackrel{(d)}{=} & n_T-n_T\delta,
  \label{eqn:binomial-ratio-telescopic-4}
\end{IEEEeqnarray}
where in $(a)$ we used that $\Pr[G \ge j]\geq \Pr[G = j]$;
in $(b)$ we used that $1-\Pr[G \ge j] = \Pr[G < j]$;
$(c)$ follows by reindexing, since each $\Pr[G=k]$ appears exactly $n_T-k$ times in the sum;
and $(d)$ follows since $\sum_{k=0}^{n_T} \Pr[G=k] = 1$  and $\sum_{k=0}^{n_T} \Pr[G=k] k = \mathbb{E}[G] = n_T\delta$.

Using \eqref{eqn:binomial-ratio-telescopic-4} in
\eqref{eqn:MI_almost_fin_cor}, combining the resulting expression with \eqref{eqn:IXY_J-1}--\eqref{eqn:IXY_J-3}, and taking the limit as $n\to\infty$, we obtain the desired upper bound \eqref{eqn:upper-bound-cor} on the exchangeable capacity.

\section{Proof of Proposition~\ref{Lozano_bursty}} \label{Proof_Lozano_bursty}
Let
\begin{equation}
X_{u,k} = B_{u} \tilde{X}_{u,k}, \quad u=1,\ldots,n_T,
\end{equation}
where $B_u\sim\text{Ber}(\delta)$ and $\tilde{X}_{u,k}$ is independent of $\mathsf{P}$, has unit energy, and is non-zero almost surely. Let $\vect{X}_k = [X_{1,k},\ldots,X_{n_T,k}]^T$, $\vect{B}=[B_1,\ldots,B_{n_T}]^T$, and $\tilde{\vect{X}}_{k}=[\tilde{X}_{1,k},\ldots,\tilde{X}_{n_T,k}]^T$. The mutual information can then be upper-bounded as
\begin{IEEEeqnarray}{lCl}
    I(\vect{X}_k;\vect{Y}_k) &\stackrel{(a)}{\leq}& I({\vect{X}_k};\vect{Y}_k,{\vect{B}})\nonumber\\
    &\stackrel{(b)}{=}& I(\vect{X}_k;{\vect{B}})+I({\vect{X}_k};\vect{Y}_k|{\vect{B}})\nonumber\\
    &\stackrel{(c)}{\leq}& n_T H_b(\delta)+ I(\vmat{H}_k,\vect{X}_k;\vect{Y}_k|{\vect{B}})-I(\vect{Y}_k;\vmat{H}_k|\vect{X}_k,{\vect{B}}),\label{Lozano_I0}
\end{IEEEeqnarray}
where $(a)$ follows by giving the extra information $\vect{B}$;  $(b)$ follows from the chain rule of mutual information; $(c)$ follows by upper-bounding $I(\vect{X}_k;{\vect{B}})$ by the entropy of the independent binary random variables $B_1,\ldots,B_{n_T}$ and by using the chain rule of mutual information to add and subtract the extra information $\vmat{H}_k$.

We continue by bounding the first mutual information on the RHS of \eqref{Lozano_I0} as
\begin{IEEEeqnarray}{lCl}
    I(\vmat{H}_k,\vect{X}_k;\vect{Y}_k|{\vect{B}}) & = & h(\vect{Y}_k|{\vect{B}}) - h(\vect{Y}_k|\vmat{H}_k,\vect{X}_k,{\vect{B}}) \nonumber\\
    & \stackrel{(a)}{\leq} & \sum_{r=1}^{n_R} \log\left(1+ \mathsf{P}\sum_{u=1}^{n_T}g_{r,u}\right) \nonumber\\
    & \stackrel{(b)}{=} & n_R \log(1+ \mathsf{P})\label{Lozano_I1}
    \end{IEEEeqnarray}
where $(a)$ follows by upper-bounding the first entropy by that of a Gaussian random vector with the same second moment as $\vect{Y}_k$ \cite[Th.~9.6.5]{Cover}, and by noting that the second entropy is the entropy of $\vect{Z}_k$, which has a Gaussian distribution; and $(b)$ follows because, by the proposition's assumption, $g_{r,1} + \ldots + g_{r,n_T}=1$. 

We next analyze the second mutual information on the RHS of \eqref{Lozano_I0} as follows:
\begin{IEEEeqnarray}{lCl}\label{Lozano_I2}
    I(\vect{Y}_k; \vmat{H}_k|\vect{X}_k,{\vect{B}}) &\stackrel{}{=}& \sum_{r=1}^{n_R} I(Y_{r,k}; \vmat{H}_k|\vect{X}_k,{\vect{B}}) \nonumber\\
    &\stackrel{}{=}& \sum_{r=1}^{n_R} \E{\log\left(1+\mathsf{P}\sum_{u=1}^{n_T} B_t |\tilde{X}_u|^2 g_{r,u}\right)},
    \end{IEEEeqnarray}
where we used that, conditioned on $\vect{X}_k$ and $\vect{B}$, the entries in $\vect{Y}_k$ are independent of each other and have a Gaussian distribution, hence the mutual information can be evaluated in closed form.

We lower-bound the RHS of \eqref{Lozano_I2} by distinguishing between the case where $B_1=\ldots=B_{n_T} =0$ and the case where at least one $B_t$ is $1$. In the former case, the logarithm inside the expectation is zero. In the latter case, we can lower-bound it as
\begin{equation}
\log\left(1+\mathsf{P}\sum_{u=1}^{n_T} B_t |\tilde{X}_u|^2 g_{r,u}\right) \geq \log\left(\mathsf{P} |\tilde{X}_{u'}|^2 g_{r,u'}\right)
\end{equation}
for any arbitrary $u'=1,\ldots,n_T$ for which $B_{u'}=1$. It follows that
\begin{equation}
\label{Lozano_I3}
 I(\vect{Y}_k; \vmat{H}_k|\vect{X}_k,{\vect{B}}) \geq \left(1- (1-\delta)^{n_T}\right) \min_{u=1,\ldots,n_T}\E{\log\left(\mathsf{P} |\tilde{X}_{u}|^2g_{r,u}\right)}.
\end{equation}

Combining \eqref{Lozano_I1} and \eqref{Lozano_I3} with \eqref{Lozano_I0}, we obtain the upper bound
\begin{IEEEeqnarray}{lCl}
I(\vect{X}_k;\vect{Y}_k) & \leq & n_R(1-\delta)^{n_T}\log\mathsf{P} + n_T H_b(\delta)+ n_R\log\left(1 + \frac{1}{\mathsf{P}}\right) - \left(1- (1-\delta)^{n_T}\right)\sum_{r=1}^{n_R} \E{\log(|\tilde{X}_{u_{\star}}|^2g_{r,u_{\star}})},
\end{IEEEeqnarray}
where
\begin{equation}
u_{\star} \eqdef \arg \min_{u=1,\ldots,n_T} \E{\log\left(|\tilde{X}_{u}|^2g_{r,u}\right)}.
\end{equation}
This is \eqref{eq:Lozano_bursty} in Proposition~\ref{Lozano_bursty}.

\section{Proof of Lemma~\ref{lem:double-exponential-decay}}\label{apx:double-exponential-decay}

By \eqref{eqn:double-exponential-decay-alpha}, we have
\begin{equation}
\label{eqn:double-exponential-decay-0.5}
\sum_{\ell=L_{\mathsf{P}}+1}^{\infty} \alpha_{\ell} \leq \sum_{\ell=L_{\mathsf{P}}+1}^{\infty} \frac{1}{\exp(\exp(\ell^{a}))}.
\end{equation}
Any $L_{\mathsf{P}}$ satisfying
\begin{equation}\label{eqn:double-exponential-decay-1}
    \sum_{\ell=L_{\mathsf{P}}+1}^{\infty} \frac{1}{\exp(\exp(\ell^{a}))} \leq \frac{\sigma^2}{\mathsf{P}}
\end{equation}
will therefore also satisfy \eqref{eqn:conditionL}. To find such an $L_{\mathsf{P}}$, we first rewrite the sum in \eqref{eqn:double-exponential-decay-1} as
\begin{equation}\label{eqn:double-exponential-decay-2}
    \sum_{\ell=L_{\mathsf{P}}+1}^{\infty} \frac{1}{\exp(\exp(\ell^{a}))} 
     = \frac{1}{\exp\bigl(\exp\bigl(L_{\mathsf{P}}^{a}\bigr)\bigl)}
           \sum_{\ell=L_{\mathsf{P}}+1}^{\infty} \exp\bigl(\exp(L_{\mathsf{P}}^{a})-\exp(\ell^{a})\bigl).
\end{equation}
Since the function $x\mapsto e^{x^a}\!-x^a$, $a\geq 1$ is monotonically increasing, we next note that, for every $\ell \geq L_{\mathsf{P}}+1$, 
\begin{equation}\label{eqn:double-exponential-decay-3a}
\exp\left( \exp\bigl(L_{\mathsf{P}}^{a}\bigr) - \exp(\ell^{a}) \right)
  \leq \exp\bigl(L_{\mathsf{P}}^{a} - \ell^{a} \bigr).
\end{equation}
Using a first-order Taylor series of $\ell^{a}$ around $\ell=L_{\mathsf{P}}$,
and that $x\mapsto x^{a}$ is a convex function for $a\geq 1$, we obtain
\begin{equation}\label{eqn:double-exponential-decay-3b}
 \ell^{a} \geq L_{\mathsf{P}}^{a} + a L_{\mathsf{P}}^{a-1}\bigl(\ell-L_{\mathsf{P}}\bigr).
\end{equation}
It follows that
\begin{IEEEeqnarray}{lCl} \label{eqn:double-exponential-decay-4}
\sum_{\ell=L_{\mathsf{P}}+1}^{\infty} \exp\bigl(\exp(L_{\mathsf{P}}^{a})-\exp(\ell^{a})\bigl)
  &\leq & \sum_{\ell = L_{\mathsf{P}}+1}^{\infty} \exp\bigl(a L_{\mathsf{P}}^{a} - \ell a L_{\mathsf{P}}^{a-1}\bigr)\nonumber\\
  & = & \frac{\exp\bigl(-a L_{\mathsf{P}}^{a-1}\bigr)}{1-\exp\bigl(- a L_{\mathsf{P}}^{a-1}\bigr)} \nonumber\\
  & \leq & \frac{e^{-a}}{1-e^{-a}}
  \label{eqn:double-exponential-decay-5}
\end{IEEEeqnarray}
by the expression of the geometric sum, and because the function $x\mapsto e^{-x}/(1-e^{-x})$ is monotonically decreasing in $x$ and $L_{\mathsf{P}} \geq 1$. We thus have that
\begin{equation}
\label{eqn:double-exponential-decay-6a}
 \sum_{\ell=L_{\mathsf{P}}+1}^{\infty} \frac{1}{\exp(\exp(\ell^{a}))} \leq \frac{1}{\exp\bigl(\exp\bigl(L_{\mathsf{P}}^{a}\bigr)\bigl)} \frac{e^{-a}}{1-e^{-a}}.
\end{equation}

We finish the proof by noting that any integer $L_{\mathsf{P}}$ satisfying
\begin{equation}
L_{\mathsf{P}} \geq \left(\log\log \left(\frac{\mathsf{P}}{\sigma^2} \frac{e^{-a}}{1-e^{-a}} \right)\right)^{\frac{1}{a}}
\end{equation}
yields
\begin{equation}
\frac{1}{\exp\bigl(\exp\bigl(L_{\mathsf{P}}^{a}\bigr)\bigl)} \frac{e^{-a}}{1-e^{-a}} \leq \frac{\sigma^2}{\mathsf{P}}.
\end{equation}
By \eqref{eqn:double-exponential-decay-0.5} and \eqref{eqn:double-exponential-decay-6a}, the same $L_{\mathsf{P}}$ also satisfies \eqref{eqn:conditionL}. This proves Lemma~\ref{lem:double-exponential-decay}.

\section{Proof of Theorem~\ref{thm:lower-bound}}\label{sec:proof-lower-bound}

Consider the bursty signaling scheme introduces in Section~\ref{sec:bursty-signaling}, where the signal transmitted by user $u$ at time $k$ is given by
\begin{equation}
X_{\ell,u,k}= B_{\ell,1}\hat{B}_{\ell,u}\hat{X}_{\ell,u,k}, \quad \ell=0,1,\ldots
\end{equation}
Here, $B_{\ell,u}\sim\textnormal{Ber}(\delta)$ models the user activity, $\hat{B}_{\ell,1}\sim\textnormal{Ber}(\xi)$ is a random variable that artificially introduces burstiness, $\{\hat{X}_{\ell,1,k}\}$ are i.i.d.\ circularly-symmetric random variables with $\log|\hat{X}_{\ell,1,k}|$ uniformly distributed over the interval $[0,P]$, and $\hat{B}_{\ell,u}$ and $\{\hat{X}_{\ell,u,k}\}$ are zero for $u=2,\ldots,n_T$; cf.~\eqref{eq:trans_signal} and \eqref{eq:bursty_signaling}.

Since the only active user is $u=1$, we will omit the subscript $u$ in the remainder of this proof. To simplify notation, let $\tilde{B}_{\ell}\triangleq B_{\ell}\hat{B}_{\ell}$. Since $B_{\ell}$ and $\hat{B}_{\ell}$ are independent, $\tilde{B}_{\ell}$ has a Bernoulli distribution with activation probability $\delta\xi$. By \eqref{Capacity-red0}, a lower bound on $C(\mathsf{P})$ follows by lower-bounding the mutual information $I(\vect{X}_{0,1}^n;\vect{Y}_1^n)$ as
\begin{IEEEeqnarray}{lCl}
    I(X_{0,1}^n;\vect{Y}_1^n)&\stackrel{(a)}{=}&I(X_{0,1}^n;\tilde{B}_1^{L_{\mathsf{P}}}) + I(X_{0,1}^n;\vect{Y}_1^n | \tilde{B}_{1}^{L_{\mathsf{P}}})  - I(X_{0,1}^n;\tilde{B}_{1}^{L_{\mathsf{P}}}|\vect{Y}_1^n)\nonumber\\
    &\stackrel{(b)}{=}& I(X_{0,1}^n;\tilde{B}_1^{L_{\mathsf{P}}}) + I(X_{0,1}^n;\vect{Y}_1^n |\tilde{B}_{1}^{L_{\mathsf{P}}}) - H(\tilde{B}_{1}^{L_{\mathsf{P}}}|\vect{Y}_1^n) + H(\tilde{B}_{1}^{L_{\mathsf{P}}}|\vect{Y}_1^n,X_{0,1}^n) \nonumber\\
    &\stackrel{(c)}{\geq} & I(X_{0,1}^n;\vect{Y}_1^n | \tilde{B}_{1}^{L_{\mathsf{P}}}) - L_{\mathsf{P}}\nonumber\\
    &\stackrel{(d)}{\geq} & I(\hat{X}_{0,1}^n, \tilde{B}_0;\vect{Y}_1^n|\tilde{B}_{1}^{L_{\mathsf{P}}})-L_{\mathsf{P}} \nonumber\\
    &\stackrel{(e)}{=}& I(\tilde{B}_0;\vect{Y}_1^n|\tilde{B}_{1}^{L_{\mathsf{P}}}) + I(\hat{X}_{0,1};\vect{Y}_1^n|\tilde{B}_{1}^{L_{\mathsf{P}}}, \tilde{B}_0) -L_{\mathsf{P}} \nonumber\\
    &\stackrel{(f)}{\geq}& I(\hat{X}_{0,1}^n;\vect{Y}_1^n|\tilde{B}_{1}^{L_{\mathsf{P}}},\tilde{B}_0) -L_{\mathsf{P}}, \label{Imain_bursty}
\end{IEEEeqnarray}
where $L_{\mathsf{P}}$ is an integer satisfying \eqref{eqn:conditionL}. Here, $(a)$ follows from the chain rule for mutual information \cite[Th.~2.5.2]{Cover}; $(b)$ follows by expressing the last mutual information as a difference of conditional entropies of the discrete random variables $\tilde{B}_{1}^{L_{\mathsf{P}}}$; $(c)$ follows by the nonnegativity of mutual information and entropy \cite[Lemma~2.1.1 \& Eq.~(2.90)]{Cover}, and by upper-bounding the entropy $H( \tilde{B}_{1}^{L_{\mathsf{P}}}|\vect{Y}_1^n)$ by the logarithm of the number of possible values $\tilde{B}_{1}^{L_{\mathsf{P}}}$ can take \cite[Th.~2.6.4]{Cover}; $(d)$ follows because $X_{0,1}^n$ is a function of $\hat{X}_{0,1}^n$ and $\tilde{B}_0$ and by the data processing inequality \cite[Th.~2.8.1]{Cover}; $(e)$ follows again from the chain rule of mutual information; $(f)$ follows again from the nonnegativity of mutual information.

To lower-bound the mutual information on the RHS of \eqref{Imain_bursty}, we write the mutual information $I(\hat{X}_{0,1}^n;\vect{Y}_1^n|\tilde{B}_{1}^{L_{\mathsf{P}}},\tilde{B}_0)$ as an average of the mutual informations $I(\hat{X}_{0,1}^n;\vect{Y}_1^n|\tilde{B}_{1}^{L_{\mathsf{P}}}=\tilde{b}_{1}^{L_{\mathsf{P}}},\tilde{B}_0=\tilde{b}_0)$, and lower-bound these mutual informations as follows: If $\tilde{b}_0 = 0$ or $\tilde{b}_{1}^{L_{\mathsf{P}}}\neq\vect{0}$, then we use the nonnegativity of mutual information to bound
\begin{equation}
\label{eq:appB_MI_0}
    I(\hat{X}_{0,1}^n;\vect{Y}_1^n|\tilde{B}_{1}^{L_{\mathsf{P}}}=\tilde{b}_{1}^{L_{\mathsf{P}}},\tilde{B}_0=\tilde{b}_0) \geq 0.
\end{equation}
If $\tilde{b}_0=1$ and $\tilde{b}_{1}^{L_{\mathsf{P}}}=\vect{0}$, then we use the following lemma:
\begin{lemma}\label{lem:single-user-lowerbound}
    We have that
    \begin{IEEEeqnarray}{lCl}
I(\hat{X}_{0,1}^n;\vect{Y}_1^n|\tilde{B}_{1}^{L_{\mathsf{P}}}=\vect{0},\tilde{B}_0=1) & \geq &n\left(\log\log\mathsf{P} -\gamma - \log(e) -2\log\left(1+\sqrt{2}\sigma\right)\right)\nonumber\\
& \triangleq & n \underline{R}(\mathsf{P}).\label{Ilemma_bursty}
\end{IEEEeqnarray}
\end{lemma}
\begin{IEEEproof}
The result is a lower bound on the capacity of a noncoherent point-to-point fading channel with non-Gaussian additive noise and is based on a lower bound presented in \cite[Lemma~4]{Lapidoth2005}. For completeness, we provide the proof in the subsection at the end of this appendix.
\end{IEEEproof}
It follows from \eqref{eq:appB_MI_0} and \eqref{Ilemma_bursty} that
\begin{IEEEeqnarray}{lCl}
I(\hat{X}_{0,1}^n;\vect{Y}_1^n|\tilde{B}_{1}^{L_{\mathsf{P}}},\tilde{B}_0) & \geq & n\delta\xi(1-\delta\xi)^{L_{\mathsf{P}}} \underline{R}(\mathsf{P}),\label{I_bursty_3}
\end{IEEEeqnarray}
where we also used that $\Pr(\tilde{B}_0=1)=\delta\xi$ and
\begin{equation}
    \Pr\left(\tilde{B}_{1}^{L_{\mathsf{P}}}=\vect{0}\right) = (1-\delta\xi)^{L_{\mathsf{P}}}.
\end{equation}
Replacing \eqref{I_bursty_3} in \eqref{Imain_bursty}, and dividing both sides of the inequality by $n$, we finally obtain
\begin{IEEEeqnarray}{lCl}
\frac{1}{n}I(\vect{X}_{0,1}^n;\vect{Y}_1^n) 
&\geq& \delta\xi(1-\delta\xi)^{L_{\mathsf{P}}} \underline{R}(\mathsf{P}) -\frac{1}{n} L_{\mathsf{P}}.\label{I13_bursty}
\end{IEEEeqnarray}
The lower bound \eqref{eqn:lower-bound} in Theorem~\ref{thm:lower-bound} follows then from \eqref{I13_bursty} and \eqref{Capacity-red0} upon letting $n$ tend to infinity.

\subsection{Proof of Lemma~\ref{lem:single-user-lowerbound}}
\label{app:lem_single-user-lowerbound}
It follows from the chain rule for mutual information that
\begin{IEEEeqnarray}{lCl}
    I(\hat{X}_{0,1}^n;\vect{Y}_1^n|\tilde{B}_{1}^{L_{\mathsf{P}}}=\vect{0},\tilde{B}_0=1) &=& \sum_{k=1}^n I(\hat{X}_{0,k};\vect{Y}_1^n|\tilde{B}_{1}^{L_{\mathsf{P}}}=\vect{0},\tilde{B}_0=1,\hat{X}_{0,1}^{k-1})\nonumber\\
    &\geq & \sum_{k=1}^n I(\hat{X}_{0,k};\vect{Y}_k|\tilde{B}_{1}^{L_{\mathsf{P}}}=\vect{0},\tilde{B}_0=1),\label{I0_bursty}
\end{IEEEeqnarray}
where we used in the second step that the channel inputs are independent and that removing the channel outputs $\vect{Y}_{\ell}$, $\ell \neq k$ reduces mutual information.

We next use the assumptions $\tilde{B}_0=1$ and $\vect{\tilde{B}}_{1}^{L_{\mathsf{P}}}=\vect{0}$ to express the channel output $\vect{Y}_k$ as
\begin{equation}
\vect{Y}_k=\vect{H}_{0,k}\hat{X}_{0,k}+\vect{W}_k,
\end{equation}
where
\begin{equation}
    \vect{W}_k \triangleq \sum_{\ell=L_{\mathsf{P}}+1}^{\infty} \vect{H}_{\ell,k}X_{\ell,k}+\vect{Z}_k
\end{equation}
and $\vect{H}_{\ell,k}$ are $(n_R \times 1)$-vectors containing the fading coefficients from user $1$ in cell $\ell$ to the receive antennas in the intended cells.

With these definitions, we can lower-bound the mutual information in \eqref{I0_bursty} as
\begin{IEEEeqnarray}{lCl}
I(\hat{X}_{0,k};\vect{Y}_k|\tilde{B}_{1}^{L_{\mathsf{P}}}=\vect{0},\tilde{B}_0=1) & = & I(\hat{X}_{0,k};\vect{H}_{0,k}\hat{X}_{\vect{\tilde{b}}_0,k}+\vect{W}_k) \nonumber\\
& \geq & I(\hat{X}_{0,k};H_{0,1,k}\hat{X}_{0,k}+W_k), \label{eq:LB_ent}
\end{IEEEeqnarray}
where $H_{0,1,k}$ denotes the time-$k$ fading coefficient from user $1$ to receive antenna $1$, and $W_k$ denotes the first component of $\vect{W}_k$. Here, the inequality follows by ignoring the received signal at receive antennas $2,\ldots,n_R$.

We next lower-bound $I(\hat{X}_{0,k};H_{0,1,k}\hat{X}_{0,k}+W_k)$ using the following lemma:

\begin{lemma}[Lapidoth 2005]
\label{lemma:lapidoth05}
Let the random variables $X$, $H$, and $W$ have finite second moments. Assume that both $X$ and $H$ are of finite differential entropy. Finally, assume that $X$ is independent of $H$; that $X$ is independent of $W$; and that $X\markov H \markov W$ forms a Markov chain. Then,
\begin{equation}
I(X;HX + W) \geq h(X) - \E{\log|X|^2} + \E{\log|H|^2} - \E{\log\left(\pi e\left(\sigma_H + \frac{\sigma_W}{|X|}\right)^2\right)},
\end{equation}
where $\sigma_H>0$ and $\sigma_W\geq 0$ denotes the standard deviations of $H$ and $W$, respectively. 
\end{lemma}
\begin{IEEEproof}
See \cite[Lemma~4]{Lapidoth2005}.
\end{IEEEproof}

It is easy to check that $X=\hat{X}_{0,k}$, $H=H_{0,1,k}$, and $W=W_k$ satisfy the lemma's condition. In particular, by the definition of $L_{\mathsf{P}}$ and our choice of $X_{\ell,k}$,
\begin{equation}
\label{eq:UB_Wk}
\E{|W_k|^2} \leq \sum_{\ell=L_{\mathsf{P}}+1}^{\infty} \alpha_{\ell}\mathsf{P} + \sigma^2 \leq 2\sigma^2.
\end{equation}
It thus follows from Lemma~\ref{lemma:lapidoth05} that
\begin{IEEEeqnarray}{lCl}
I(\hat{X}_{0,k};H_{0,1,k}\hat{X}_{0,k}+W_k) & \geq & h(\hat{X}_{0,k}) - \E{\log|\hat{X}_{0,k}|^2} + \E{\log|H_{0,1,k}|^2} - \log(\pi e) - 2\E{\log\left(1+\frac{\sqrt{\E{|W_k|^2}}}{|\hat{X}_{0,k}|}\right)} \nonumber\\
& \geq & h(\hat{X}_{0,k}) - \E{\log|\hat{X}_{0,k}|^2} + \E{\log|H_{0,1,k}|^2} - \log(\pi e) - 2\log\left(1+\sqrt{2}\sigma\right),\label{LB_h+1}
\end{IEEEeqnarray}
where the second inequality follows from \eqref{eq:UB_Wk} and because $|\hat{X}_{0,k}|\geq 1$.

The differential entropy on the RHS of \eqref{LB_h+1} can be evaluated as
\begin{IEEEeqnarray}{lCl}
h(\hat{X}_{0,k}) & = & h(\log|\hat{X}_{0,k}|^2) + \E{\log |\hat{X}_{0,k}|^2} + \log \pi \nonumber\\
& = & \log\log\mathsf{P} + \E{\log |\hat{X}_{0,k}|^2} + \log \pi, \label{hX}
\end{IEEEeqnarray}
where the first step follows expressing the differential entropy of a circularly-symmetric random variable by the differential entropy of the logarithm of its absolute value \cite[Eqs. (326) and (316)]{lapidoth2003capacity}; and the second step follows by computing the differential entropy of the uniformly-distributed random variable $\log|\hat{X}_{0,k}|^2$.

Similarly, the expected value of the logarithm of the exponentially-distributed random variable $|H_{0,k}|^2$ can be computed as (see, e.g.,\cite[Eqs. (209)-(212)]{lapidoth2003capacity})
\begin{equation}
\E{\log|H_{0,1,k}|^2} = -\gamma. \label{ElogH}
\end{equation}
Combining \eqref{hX} and \eqref{ElogH} with \eqref{LB_h+1}, we obtain
\begin{equation}
I(\hat{X}_{0,k};H_{0,1,k}\hat{X}_{0,k}+W_k) \geq \log\log\mathsf{P} -\gamma - \log(e) -2\log\left(1+\sqrt{2}\sigma\right).
\end{equation}
Together with \eqref{eq:LB_ent} and \eqref{I0_bursty}, this yields
\begin{equation}
 I(\hat{X}_{0,1}^n;\vect{Y}_1^n|\tilde{B}_{1}^{L_{\mathsf{P}}}=\vect{0},\tilde{B}_0=1) \geq n\left(\log\log\mathsf{P} -\gamma - \log(e) -2\log\left(1+\sqrt{2}\sigma\right)\right)
\end{equation}
which proves Lemma \ref{lem:single-user-lowerbound}.

\section{Proof of Corollary~\ref{cor:on-off-signaling-lower-bound}}\label{apx:on-off-signaling-lower-bound}

By assumption, the fading variances $\alpha_\ell$ satisfy \eqref{eqn:double-exponential-decay-alpha} for some $a > 1$. It thus follows from Lemma~\ref{lem:double-exponential-decay} that any integer $L_{\mathsf{P}}$ satisfying
\begin{equation}
\label{eq:LB_LP}
L_{\mathsf{P}} > \left(\log\log \left(\frac{\mathsf{P}}{\sigma^2} \frac{e^{-a}}{1-e^{-a}} \right)\right)^{\frac{1}{a}}
\end{equation}
also satisfies \eqref{eqn:conditionL}. We next choose the activation probability
\begin{equation}
\xi_{\mathsf{P}} = \bigl( \log \log \mathsf{P} \bigr)^{-\frac{1+\varepsilon}{a}}
\end{equation}
for some arbitrary $0<\varepsilon<a-1$. We then analyze the leading term in \eqref{eqn:lower-bound}, given by,
\begin{equation}
\label{eqn:on-off-signalling-0}
\delta\xi_{\mathsf{P}}(1-\delta\xi_{\mathsf{P}})^{L_{\mathsf{P}}}\log\log\mathsf{P}
\end{equation}
asymptotically as $\mathsf{P}\to\infty$. To this end, we first note that, if
\begin{equation}
\label{eqn:on-off-signalling-0-assumption}
\varliminf_{\mathsf{P}\to\infty} (1-\delta\xi_{\mathsf{P}})^{L_{\mathsf{P}}} > 0
\end{equation}
then there exist positive constants $\nu$ and $\mathsf{P}_0$ such that, for $\mathsf{P} \geq \mathsf{P}_0$, the leading term in \eqref{eqn:lower-bound} is lower-bounded as
\begin{equation}
\delta\xi_{\mathsf{P}}(1-\delta\xi_{\mathsf{P}})^{L_{\mathsf{P}}}\log\log\mathsf{P} \geq \delta\xi_{\mathsf{P}} \nu \log\log\mathsf{P} = \delta\nu \left(\log\log\mathsf{P}\right)^{1-\frac{1+\varepsilon}{a}}
\end{equation}
from which Corollary~\ref{cor:on-off-signaling-lower-bound} follows.

It thus remains to prove \eqref{eqn:on-off-signalling-0-assumption}. To this end, we consider the logarithm of $(1-\delta\xi_{\mathsf{P}})^{L_{\mathsf{P}}}$ and show that it is finite. Indeed, we have that
\begin{IEEEeqnarray}{lCl}
{L_{\mathsf{P}}} \log(1-\delta\gamma_{\mathsf{P}})
    & > &  \left(\log\log \left(\frac{\mathsf{P}}{\sigma^2} \frac{e^{-a}}{1-e^{-a}} \right)\right)^{\frac{1}{a}}\log\left(1-\frac{\delta}{\bigl( \log \log \mathsf{P} \bigr)^{\frac{1+\varepsilon}{a}}}\right)\\
    &\geq & -  \left(\log\log \left(\frac{\mathsf{P}}{\sigma^2} \frac{e^{-a}}{1-e^{-a}} \right)\right)^{\frac{1}{a}}
            \frac{\delta}{\bigl(\log \log \mathsf{P} \bigr)^{\frac{1+\epsilon}{a}}}
       \left(1-\frac{\delta}{\bigl( \log \log \mathsf{P} \bigr)^{\frac{1+\epsilon}{a}}}\right)^{-1}, \label{eqn:on-off-signalling-1a}
\end{IEEEeqnarray}
where in the second step we used that for $0\leq z < 1$, it holds $\log(1-z) \geq -\frac{z}{1-z}$.

The first term on the RHS of \eqref{eqn:on-off-signalling-1a} is of order $(\log\log\mathsf{P})^{1/a}$; the second term is of order $(\log\log\mathsf{P})^{-(1+\varepsilon)/a}$; the third term tends to $1$ as $\mathsf{P}\to\infty$. It follows that the RHS of \eqref{eqn:on-off-signalling-1a} is of order $(\log\log\mathsf{P})^{-\varepsilon/a}$, which tends to zero as $\mathsf{P}\to\infty$. Since the limit inferior of the logarithm of $(1-\delta\xi_{\mathsf{P}})^{L_{\mathsf{P}}}$ is lower-bounded by zero, we conclude that
\begin{equation}
\varliminf_{\mathsf{P}\to\infty} (1-\delta\xi_{\mathsf{P}})^{L_{\mathsf{P}}} \geq 1.
\end{equation}
In fact, we have $(1-\delta\xi_{\mathsf{P}})^{L_{\mathsf{P}}}\leq 1$ for every $\mathsf{P}$, so it even holds that $(1-\delta\xi_{\mathsf{P}})^{L_{\mathsf{P}}}\to 1$ as $\mathsf{P}\to\infty$. This proves \eqref{eqn:on-off-signalling-0-assumption} and concludes the proof of Corollary~\ref{cor:on-off-signaling-lower-bound}.

\ifCLASSOPTIONcaptionsoff
  \newpage
\fi

\end{document}